\documentclass[pra,twocolumn]{revtex4}

\usepackage{graphics}
\usepackage{graphicx}
\usepackage{bm}
\usepackage{latexsym}
\usepackage{color}
\usepackage{mathrsfs}
\usepackage{braket}
\usepackage{enumerate}
\usepackage[usenames,dvipsnames]{xcolor}
\usepackage[colorlinks=true,citecolor=blue,linkcolor=blue,urlcolor=blue,backref=true]{hyperref}
\usepackage{float}
\usepackage[raggedright,tight]{subfigure}  
\usepackage{amsmath}
\usepackage{amsfonts}
\usepackage{braket}
\usepackage[dvipsnames]{xcolor}
\usepackage{rotating}
\usepackage{multirow}
\usepackage{xcolor}
\usepackage[normalem]{ulem}

\begin{document}

\title{Non-conjugate quantum subsystems}
\author{Adam Stokes}\email{adamstokes8@gmail.com}
\affiliation{School of Mathematics, Statistics, and Physics, Newcastle University, Newcastle upon Tyne, Tyne and Wear, NE1 7RU, United Kingdom}

\date{\today}

\begin{abstract}
We introduce an alternative way to understand the decomposition of a quantum system into interacting parts and show that it is natural in several physical models. This enables us to define a reduced density operator for a working system interacting with a thermal bath that is consistent with the inclusion of the interaction Hamiltonian within the working system's energy. We subsequently provide a self-consistent formulation of quantum thermodynamics that incurs non-trivial physical corrections to thermodynamic relations and quantities previously defined within the literature.

\end{abstract}


\maketitle

\section{Introduction}

Physical processes are understood in terms of interactions between physical systems, so defining the notion of {\em interacting subsystem} can be considered a foremost task in theoretical physics. 
In a statistical theory, the probabilities $P(o_A)$ and $P(o_B)$ to obtain the outcomes $o_A$ and $o_B$ in measurements performed on statistically independent subsystems $A$ and $B$ should be those of independent events; $P(o_A \cap o_B)=P(o_A)P(o_B)$, otherwise the subsystems are correlated. This requirement is met in quantum theory, because the inner-product on a tensor-product space ${\cal H}_A\otimes {\cal H}_B$ is defined in terms of the constituent inner-products by $(\bra{\psi_A}\otimes \bra{\psi_B})(\ket{\phi_A}\otimes \ket{\phi_B})=\braket{\psi_A|\phi_A}\braket{\psi_B|\phi_B}$. 
More generally, subsystem density operators are defined using the partial trace by $\rho_A={\rm tr}_B\rho$ and $\rho_B={\rm tr}_A\rho$, where $\rho$ is the density operator representing the generally mixed state of the composite. By definition, the {\em state} of a systems must suffice to provide complete statistical predictions pertaining to any system {\em observable}. Observables of subsystems $A$ and $B$ must therefore be represented by operators of the form $O_A\equiv O_A\otimes I_B$ and $O_B\equiv I_A\otimes O_B$ respectively, where $I_A$ and $I_B$ are the identity operators on ${\cal H}_A$ and ${\cal H}_B$. We then have that ${\rm tr}(O_s\rho)={\rm tr}(O_s\rho_s),~s=A,\,B$.

This framework is obviously sufficient whenever interactions are not present, but whether it is able to fully capture all of the properties expected of {\em interacting} subsystems is less clear. Specifically, according to these definitions, even when subsystems $A$ and $B$ are interacting, such that they cannot be spatially separated, their observables are trivially compatible, so an operation such as a measurement performed on subsystem $A$ cannot disturb subsystem $B$. Quantum theory dictates however, that a measurement of $A$ immediately disturbs $A$, and if by definition of an {\em interaction}, system $A$ is able to immediately disturb $B$, then a measurement of $A$ must be able to disturb $B$. A fundamental example is provided by quantum electrodynamics based on Maxwell's equations; if system $A$ is a charged particle and system $B$ is the surrounding transverse electromagnetic field, then to obtain the correct Maxwell-Lorentz equations it must be the case that the transverse electric field is not compatible with the charged particle's energy-momentum \cite{cohen-tannoudji_photons_1989,craig_molecular_1998,stokes_implications_2021}. 
Motivated by these observations, we introduce the concept of non-conjugate subsystems, defined as a pair of subsystems obtained using distinct Hilbert space frames rather than only one. 
Non-conjugate subsystems are allowed to possess incompatible observables and as a result they are able to exhibit fundamentally different properties to standard quantum subsystem pairs. This includes the possibility of immediately disturbing a subsystem by measuring another. We provide models in which physical subsystems are naturally non-conjugate.

We use the notion of non-conjugate subsystems to rectify an internal inconsistency occurring in formulations of quantum thermodynamics that include the system-bath interaction Hamiltonian within the definition of the working system's internal energy \cite{esposito_entropy_2010,esposito_second_2011,deffner_information_2013,strasberg_quantum_2017,strasberg_first_2021}. According to this definition, the working system's standard reduced density operator 
does not suffice to determine the average system energy and so cannot be said to represent the {\em state} of the working system. It is possible to eliminate this inconsistency by treating the working system and the bath as non-conjugate subsystems. The resulting quantum thermodynamics differs significantly from other formulations within the literature, in that it modifies the naive definition of the working system's state, rather than immediately modifying naive thermodynamic quantities \cite{seifert_first_2016,jarzynski_stochastic_2017,miller_hamiltonian_2018,strasberg_non-markovianity_2019}. However, as an emergent result, we obtain non-trivial and purely quantum corrections to previous results. The idea of non-conjugate subsystems should possess a number of further applications as a fundamental tool for understanding quantum interactions.

%

%
 
\section{Conjugate and non-conjugate quantum subsystems}\label{II}

\subsection{Conjugate quantum subsystems}

A quantum system can be defined as a pair $({\cal D},{\cal A})$ where ${\cal D}$ is the space of density operators over a Hilbert space ${\cal H}$ and ${\cal A} \supset {\cal D}$ is the algebra of Hermitian operators over ${\cal H}$. A state of the system is represented by $\rho \in {\cal D}$ and a physical observable is represented by $O\in {\cal A}$. 
A composite quantum system consisting of quantum systems $A$ and $B$ is defined using the tensor-product as $({\cal D}={\cal  D}_A\otimes {\cal D}_B,{\cal A}={\cal A}_A\otimes {\cal A}_B)$. The state of subsystem $A$ ($B$) is represented by $\rho_A ={\rm tr}_B\rho$ ($\rho_B ={\rm tr}_A\rho$). To motivate these definitions, one notes that \cite{ghirardi_general_1980}:
\begin{enumerate}[i)]
\item{Spacelike separated systems, $A$ and $B$, cannot be interacting and must possess observables represented by operators $O_A$ and $O_B$ that are trivially compatible, $[O_A,O_B]\equiv 0$, such that an operation performed on $A$ ($B$)  cannot affect $B$ ($A$).}
\end{enumerate}
An operation on system $A$ may be defined as a completely positive trace preserving map $\Phi_A$ over ${\cal D}_A$. Such a map satisfies ${\rm tr}((\Phi_A\otimes I_B[\rho])I_A\otimes O_B) = {\rm tr}(\rho I_A\otimes O_B) = {\rm tr}(\rho_BO_B)$. These equalities show that $\rho_A$ ($\rho_B$) is sufficient to provide any statistical prediction pertaining to system $A$ ($B$), so $A$ and $B$ are themselves quantum systems $({\cal D}_A,{\cal A}_A)$ and $({\cal D}_B,{\cal A}_B)$, and that it is indeed the case that an operation performed on a subsystem cannot affect the other subsystem. Note however, that these definitions do not make reference to whether or not the subsystems are interacting, and so they constitute a significantly stronger set of restrictions for defining physical subsystems than is necessitated by condition i).

\subsection{Interaction via non-commutativity}\label{noncom}

We demonstrate by way of fundamental example, that there are situations in which natural physical subsystems possess incompatible observables, such that they cannot be conventional quantum subsystems. Consider a non-relativistic atom consisting of a dynamical charge $q$ with mass $m$ and position ${\bf r}$ bound to a stationary charge $-q$ at the origin. The charge density is ${\rho}({\bf x})=q[\delta({\bf x}-{\bf r})-\delta({\bf x})]$. The momentum conjugate to ${\bf r}$ is ${\bf p}$, such that $[r_i,p_j]=i\delta_{ij}$. The eigenkets, $\{\ket{\bf p}\}$, of ${\bf p}$ such that $\braket{{\bf p}|{\bf p}'}=\delta({\bf p}-{\bf p}')$  span an abstract space ${\cal H}_A$ of a quantum system that we label $A$. An arbitrary state of $A$ can be represented using a vector $\ket{\psi} = \int d^3p\, \psi({\bf p})\ket{\bf p}$ where $\psi({\bf p}) \in L^2({\mathbb R}^3;{\mathbb C})$ is a complex square-integrable wave-function; $L^2({\mathbb R}^3;{\mathbb C}) = \{\psi: \int d^3p |\psi({\bf p})|^2 <\infty\}$. Hermitian operators of the form $f({\bf r},{\bf p})$ for some function $f$, belong to the algebra ${\cal A}_A$ of Hermitian operators over ${\cal H}_A$.

The electromagnetic field coordinate is the gauge-invariant transverse vector potential ${\bf A}_{\rm T}$. Its canonically conjugate momentum is ${\bf \Pi}$, such that $[A_{{\rm T},i}({\bf x}),\Pi_{j}({\bf x}')]=i\delta_{ij}^{\rm T}({\bf x}-{\bf x}')$. Assuming periodic boundary conditions for the volume $v$ of the quantised field allows us to write
\begin{align}
{\bf A}_{\rm T}({\bf x}) &= \sum_{{\bf k}\lambda} {{\bf e}_{{\bf k}\lambda} \over \sqrt{2\omega_k v}}\left(a^\dagger_{{\bf k}\lambda}e^{-i{\bf k}\cdot {\bf x}}+ a_{{\bf k}\lambda}e^{i{\bf k}\cdot {\bf x}}\right),\\
{\bf \Pi}({\bf x}) &= i\sum_{{\bf k}\lambda} {\bf e}_{{\bf k}\lambda} {\sqrt{\omega_k \over 2v}}\left(a^\dagger_{{\bf k}\lambda}e^{-i{\bf k}\cdot {\bf x}}- a_{{\bf k}\lambda}e^{i{\bf k}\cdot {\bf x}}\right)
\end{align}
where $\omega=|{\bf k}|$ and $a_{{\bf k}\lambda}:={\bf e}_{\bf k\lambda}\cdot (\omega{\tilde {\bf A}}_{\rm T{\bf k}}+i{\tilde {\bf \Pi}}_{\bf k})/\sqrt{2\omega}$ is the annihilation operator for a photon with momentum ${\bf k}$ and polarisation $\lambda=1,2$. Here a tilde is used to denote the Fourier transform, and the vectors ${\bf e}_{{\bf k}1},\,{\bf e}_{{\bf k}2}$ are mutually orthonormal polarisation vectors orthogonal to ${\bf k}$. The photon operators satisfy $[a_{{\bf k}\lambda},a_{{\bf k}'\lambda'}^\dagger]=\delta_{\lambda\lambda'}\delta_{\bf kk'}$.

The eigenvector $\ket{n_{{\bf k}\lambda}}$ of $a_{{\bf k}\lambda}^\dagger a_{{\bf k}\lambda}$ represents a state with $n$ photons in the mode ${\bf k}\lambda$. An arbitrary state within the Hilbert space ${\cal H}_{{\bf k}\lambda}$ of this mode may be written $\ket{\phi_{{\bf k}\lambda}} = \sum_{n} \phi_n\ket{n_{{\bf k}\lambda}}$ where $\phi_n \in \ell^2$ defines a complex square-summable sequence; $\ell^2:=\{\phi_n \in {\mathbb C} : \sum_n |\phi_n|^2 < \infty\}$. The Hilbert space ${\cal H}_B$ of the quantum system that we label $B$ is ${\cal H}_B = \bigotimes_{{\bf k}\lambda} {\cal H}_{{\bf k}\lambda}$. Hermitian functions of the form $f({\bf A}_{\rm T},{\bf \Pi})$ (or equivalently of the form $f\{a_{{\bf k}\lambda},a_{{\bf k}\lambda}^\dagger\}$) belong to the algebra ${\cal A}_B$ of Hermitian operators over ${\cal H}_B$. The composite Hilbert space and operator algebra are ${\cal H} = {\cal H}_A\otimes {\cal H}_B$ and ${\cal A}={\cal A}_A\otimes {\cal A}_B$. 

The mathematical construction of the theory's Hilbert space and operator algebra is complete, but in order to obtain physical predictions we must specify the physical {\em observables} and {\em states} that the canonical operators ${\bf z}=\{{\bf r}, {\bf p},{\bf A}_{\rm T}, {\bf \Pi}\}$ and the associated eigenvectors represent. To this end one begins by noting that a correct theory must yield the Maxwell-Lorentz equations, because these equations are empirically deduced. 
The non-dynamical Maxwell equations, $\nabla\cdot {\bf E}_{\rm L}=\rho$ and $\nabla \cdot {\bf B}=0$ for the longitudinal electric field ${\bf E}_{\rm L}$ and magnetic field ${\bf B}$ are satisfied identically by defining ${\bf E}_{\rm L} := \nabla(\nabla^{-2}\rho)$ and ${\bf B}:=\nabla \times {\bf A}_{\rm T}$. The transverse electric field is defined by ${\bf E}_{\rm T}=-{\bf {\dot A}}_{\rm T}$, such that Faraday's law ${\dot {\bf B}}=-\nabla \times {\bf E}_{\rm T}$ is satisfied identically. The remaining dynamical equations are the Lorentz force law $m{\ddot {\bf r}} = q({\bf E}({\bf r})+[{\dot {\bf r}}\times {\bf B}({\bf r})-{\bf B}({\bf r})\times {\dot {\bf r}}]/2)$, and the Maxwell-Ampere law ${\dot {\bf E}}_{\rm T} =\nabla\times {\bf B}-{\bf J}_{\rm T}$, where ${\bf J}_{\rm T}$ is the transverse component of the current ${\bf J}({\bf x})=q[{\dot {\bf r}}\delta({\bf x}-{\bf r})+\delta({\bf x}-{\bf r}){\dot {\bf r}}]/2$.

The only physical observables within these equations that remain to be specified as functions of the canonical operators ${\bf z}$, are ${\dot {\bf r}}$ and ${\dot {\bf A}}_{\rm T} = -{\bf E}_{\rm T}$. The Hamiltonian $H$ in conjunction with suitable such specifications must yield the Maxwell-Ampere law and Lorentz force law. The Hamiltonian required is nothing but the total energy, which is the energy of the atom, denoted $E_{A'}$, plus the energy of the transverse field, denoted $E_B$ \cite{cohen-tannoudji_photons_1989,craig_molecular_1998,stokes_implications_2021};
\begin{align}
&H=E_{A'}+E_B,\label{Ham}\\
&E_{A'}={1\over 2}m{\dot {\bf r}}^2 + U,\label{E}\\
&E_B= {1\over 2}\int_v d^3 x\,\left[{\bf E}_{\rm T}^2+{\bf B}^2\right]\label{Eb}
\end{align}
where \cite{cohen-tannoudji_photons_1989,craig_molecular_1998,stokes_implications_2021}
\begin{align}
U =  {1\over 2}\int_v d^3x\, {\bf E}_{\rm L}^2=\int_v d^3x \int_v d^3x'\, { \rho({\bf x})\rho({\bf x}')\over 8\pi |{\bf x}-{\bf x}'|}
\end{align}
is the Coulomb (material potential) energy. If we choose the Coulomb-gauge then the mechanical momentum $m{\dot {\bf r}}$ and transverse electric field ${\bf E}_{\rm T}=-{\dot {\bf A}}_{\rm T}$ appearing in Eqs.~(\ref{E}) and (\ref{Eb}) respectively, are \cite{cohen-tannoudji_photons_1989,craig_molecular_1998,stokes_implications_2021}
\begin{align}
&m{\dot {\bf r}}={\bf p}-q{\bf A}_{\rm T}({\bf r}) = -im[{\bf r},H],\label{can1}\\
&-{\dot {\bf A}}_{\rm T} = {\bf E}_{\rm T}=-{\bf \Pi} = i[{\bf A}_{\rm T},H].\label{can2}
\end{align}
where the equalities on the right-hand-sides can be verified by substituting Eqs.~(\ref{can1}) and (\ref{can2}) into Eq.~(\ref{Ham}) and making use of the canonical commutation relations. Similarly, using Eqs.~(\ref{can1}), (\ref{can2}), and (\ref{Ham}), one obtains the Maxwell-Ampere law and Lorentz Force law from the Heisenberg equations ${\dot {\bf E}}_{\rm T} = -i[{\bf E}_{\rm T},H]$ and $m{\ddot {\bf r}} = -i[m{\dot {\bf r}},H]$ respectively.

We have now specified all basic material and transverse electromagnetic observables ${\bf y} =\{{\bf r}, {\bf {\dot r}}, {\bf E}_{\rm T}, {\bf B}\}$ in terms of the canonical operators ${\bf z}=\{{\bf r}, {\bf p}, {\bf A}_{\rm T}, {\bf \Pi}\}$ and we have constructed a correct dynamical description defined as one that yields the correct equations of motion for the observables ${\bf y}$. We are therefore in a position to identify the physical meaning of the quantum subsystems $A$ and $B$ that we defined at the outset using the canonical operators ${\bf z}$. 
Since ${\bf B}=\nabla\times {\bf A}_{\rm T}$ and ${\bf E}_{\rm T}=-{\bf \Pi}$ the subsystem $B$ represents the transverse electromagnetic field system possessing observables of the form $f({\bf E}_{\rm T},{\bf B})$. In particular, $E_B$ in Eq.~(\ref{Eb}) belongs in ${\cal A}_B$ and is nothing but the following photonic energy \cite{cohen-tannoudji_photons_1989,craig_molecular_1998,stokes_implications_2021};
\begin{align}
&E_B=\sum_{{\bf k}\lambda} \omega\left(a^\dagger_{{\bf k}\lambda}a_{{\bf k}\lambda}+{1\over 2}\right).\label{Eb2}
\end{align}

For the material system, however, notice that we have used a subscript $A'$ when referring to the total atomic energy in Eq.~(\ref{E}) and that this is to be distinguished from the label $A$ for the mathematical quantum subsystem with canonical operators ${\bf r}$ and ${\bf p}$. Indeed, let us suppose that the current ${\bf J}$ appearing in Maxwell's equations and the corresponding momentum $m{\dot {\bf r}}={\bf p}-q{\bf A}_{\rm T}({\bf r})$ can be measured. These observables are not those of the quantum subsystem that we have called $A$, because only ${\bf p}$ belongs to ${\cal A}_A$ whereas ${\bf A}_{\rm T}$ does not. The operator $q{\bf A}_{\rm T}({\bf r})$ represents the component of the electromagnetic momentum that is generated by the electrostatic field of the dynamical charge $q$, viz., $q{\bf A}_{\rm T}({\bf r}) = \int d^3 x {\bf E}_{\rm L{\bf r}}({\bf x})\times {\bf B}({\bf x})=:{\bf P}_{\rm long}$ where ${\bf E}_{\rm L{\bf r}}:=-q\nabla [1/(4\pi|{\bf x}-{\bf r}|)]$ \cite{cohen-tannoudji_photons_1989,stokes_implications_2021}. 

If $A$ were taken as the relevant ``material" subsystem, then the ``material" momentum, for example, would be ${\bf p}=m{\dot {\bf r}}+{\bf P}_{\rm long}$, and the measurable momentum $m{\dot {\bf r}} = {\bf p}-{\bf P}_{\rm long}$ would have to be understood as a combination of ``material" and electromagnetic properties. This would contradict an {\em operational} identification of what constitutes the material system. Conversely, if we define the material system as possessing the observables that appear within the measured equations of motion, and we call this system $A'$, then it follows that $A'\neq A$. Subsequently, the only available physical interpretation of the quantum system that we have called $A$ is that it possesses observables that combine those of $A'$ and the transverse electromagnetic system $B$. In particular, ${\bf p}=m{\dot {\bf r}}+{\bf P}_{\rm long}\in {\cal A}_A$ is nothing but a combination of the measurable material momentum $m{\dot {\bf r}}$ of $A'$ and the observable ${\bf P}_{\rm long}$, which depends on observables of $B$.

If $A'$ and $B$ were standard quantum subsystems then it would necessarily be the case that $[E_{A'},E_B]= 0$ from which Eq.~(\ref{Ham}) would imply that they are not interacting, that is, both systems would simply evolve freely.
Indeed, it is easily verified that $[E_{A'},E_B]\neq 0$ because $[m{\dot r}_i,E_j({\bf x})] = i\delta_{ij}\rho({\bf x})\neq 0$ and $[m{\dot {\bf r}}^2/2,{\bf E}_{\rm T}({\bf x})]=i{\bf J}_{\rm T}({\bf x})\neq {\bf 0}$ \cite{cohen-tannoudji_photons_1989,stokes_implications_2021}, so the momentum and energy of a charge are not compatible with the transverse electric field at the position of the charge. This is necessary to obtain the Maxwell-Ampere law and the Lorentz-Force law from Eq.~(\ref{Ham}). Classically, a measurement of the charge can be considered not to alter its state, but quantum-mechanically measuring the charge's momentum collapses the state into an eigenstate of the momentum. According to the Maxwell-Ampere law this sudden change must change the transverse electric field. Even a non-selective measurement of the momentum $m{\dot {\bf r}}$ alters the subsequent statistical predictions pertaining to ${\bf E}_{\rm T}$. 
The important implication that follows immediately from expressing the theory in terms of the physical degrees of freedom ${\bf y}$ is that:
\begin{enumerate}[i)]\addtocounter{enumi}{1}
\item{The subsystems $A'$ and $B$ possess incompatible observables and in particular $[E_{A'},E_B]\neq 0$ describes an interaction between $A'$ and $B$.}
\end{enumerate}

This clearly contrasts the trivial compatibility specified within condition i). However, if subsystems are interacting they are not space-like separated. Imposing trivial compatibility of interacting subsystem observables is then overly restrictive, as the above example demonstrates. Below we propose an extension of the formalism for quantum subsystems in which such a restriction is not imposed.  


\subsection{Non-conjugate quantum subsystems}\label{noncon}

We have provided an example of a composite $C$ in which the measurable subsystem $A'$ and its bath $B$, are not those obtained from the standard decomposition $({\cal A}={\cal A}_A\otimes {\cal A}_B,{\cal H}={\cal H}_A\otimes {\cal H}_B)$, which instead yields the subsystems $A\neq A'$ and $B$. To provide an alternative decomposition, let us suppose that, as is usually the case, the atom $A'$ is much smaller than the resonant photonic wavelengths, such that it may be treated as a dipole at the origin \cite{cohen-tannoudji_photons_1989,stokes_implications_2021}. Eq.~(\ref{can1}) becomes $m{\dot {\bf r}} = {\bf p}-q{\bf A}_{\rm T}$ where ${\bf A}_{\rm T}:={\bf A}_{\rm T}({\bf 0})$. We can now realise the physical dipole $A'$ as a quantum subsystem of the composite $C$ by means of a unitary transformation $R:{\cal H}\to{\cal H}$ defined by
\begin{align}\label{pzw}
R = \exp\left(-iq{\bf r}\cdot {\bf A}_{\rm T}\right)
\end{align}
such that
\begin{align}
R[{\bf p}-q{\bf A}_{\rm T}]R^\dagger = {\bf p}.\label{r1}
\end{align}

The rotation can be viewed as transforming between two orthonormal frames (bases) of the abstract space ${\cal H}$. Specifically, given any two orthonormal basis vectors $\ket{o_\alpha}$ and $\ket{o_\beta}$ ($\alpha,\beta =1,\dots,\rm dim {\cal H}$), such that $\braket{o_\alpha|o_\beta}=\delta_{\alpha\beta}$ one can define new orthonormal vectors $R\ket{o_\alpha}=:\ket{o'_\alpha}$ and $R\ket{o_\beta}=:\ket{o_\beta'}$ such that $\braket{o'_\alpha|o_\beta'}=\delta_{\alpha\beta}$. We label these frames $X$ and $Y$.  Importantly, the associations {\em physical observable} ${\cal O}$ $\leftrightarrow$ {\em abstract operator} $O$ and {\em physical state} ${\cal S}$ $\leftrightarrow$ {\em abstract density operator} $\rho$ can only be made {\em relative} to a Hilbert space frame. To see this, suppose that in frame $X$ observable ${\cal O}$ is represented by operator $O$. In this frame the eigenvector $\ket{o_\alpha}$, such that $O\ket{o_\alpha}=o_\alpha\ket{o_\alpha}$, represents the pure state ${\cal S}$ in which the observable ${\cal O}$ necessarily possesses the (eigen)value $o_\alpha$. In frame $Y$, the same observable and state ${\cal O}$ and ${\cal S}$ are represented by $O'=ROR^\dagger$ and $\ket{o_\alpha'} = R\ket{o_\alpha}$ respectively, such that $O'\ket{o_\alpha'}=o_\alpha\ket{o_\alpha'}$. The unitarity of $R$ ensures that the eigenvalues $\{o_\alpha\}$, which are the possible outcomes of measurements of observable ${\cal O}$, are the same in both frames; $o_\alpha =\bra{o_\alpha}O\ket{o_\alpha} = \bra{o_\alpha'}O'\ket{o_\alpha'}$. More generally, if with respect to frame $X$ an observable ${\cal O}$ is represented at time $t$ by operator $O(t)$ and a state ${\cal S}$ is represented by a density operator $\rho(t)$, then with respect to frame $Y$ the same observable ${\cal O}$ and state ${\cal S}$ are represented by the different operators $O'(t)=RO(t)R^\dagger$ and $\rho'(t)=R\rho(t)R^\dagger$ respectively. Any physical prediction $\langle {\cal O}\rangle_{\cal S} = {\rm tr}[\rho(t)O(t)]= {\rm tr}[\rho'(t)O'(t)]$ for the average of an arbitrary {\em observable} ${\cal O}$ in an arbitrary {\em state} ${\cal S}$ at time $t$ is unique (frame-independent).

Physical predictions are unique, but the physical meanings of operators and vectors are {\em relative}; they are generally different in each different frame. In frame $X$ the physical dipolar momentum {\em observable} $m{\dot{\bf  r}}$ is represented by the operator ${\bf p}-q{\bf A}_T \not\in{\cal A}_A$ whereas according to Eq.~(\ref{r1}) the same observable is represented in frame $Y$ by the operator ${\bf p}\in {\cal A}_A$. Unlike in frame $X$, in frame $Y$ we have ${\bf p}=m{\dot {\bf r}}$, and so Hermitian operators of the form $f({\bf r},{\bf p})\in {\cal A}_A$ for some function $f$, can be taken as representing physical dipole observables. For example, the energy $E_{A'}$ is represented in frame $Y$ by the operator $H_A\in {\cal A}_A$, where $H_A:={\bf p}^2/(2m)+U$. The physical dipole's {\em state}, meanwhile, must by definition suffice to provide complete statistical predictions pertaining to any of the physical dipole's observables. If we let $\rho(t)$ represent the state of $C$ in frame $X$ then the density operator $\rho_A(t)={\rm tr}_B\rho(t)$ represents a state of the subsystem called $A$. The state of the physical dipole $A'$ meanwhile, is represented by the density operator $\rho_{A'}(t)={\rm tr}_B\rho'(t)$ where $\rho'(t)=R\rho(t)R^\dagger$.

Turning our attention to the bath, in frame $X$ the operator ${\bf \Pi}\in{\cal A}_B$ represents the observable $-{\bf E}_{\rm T}$. In frame $Y$ this same observable is represented by the different operator $R{\bf \Pi} ({\bf x})R^\dagger = {\bf \Pi}({\bf x}) + q{\bf r}\cdot \delta^{\rm T}({\bf x}) \not \in {\cal A}_B$, such that the same operator ${\bf \Pi}({\bf x}) \in{\cal A}_B$ represents the different observable $-{\bf E}_{\rm T}({\bf x})-q{\bf r}\cdot \delta^{\rm T}({\bf x})$. Thus, unlike in frame $X$, in frame $Y$, bath observables of the form $f({\bf E}_{\rm T},{\bf B})$ are not represented by operators belonging in ${\cal A}_B$ and Hermitian operators of the form $f({\bf A}_{\rm T},{\bf \Pi}) \in {\cal A}_B$ represent observables of some other system that we label $B'$. The physical interpretation of $B'$ is that it possesses observables which combine transverse electromagnetic and material properties, such as ${\bf \Pi}({\bf x})=-{\bf E}_{\rm T}({\bf x})-q{\bf r}\cdot \delta^{\rm T}({\bf x}) \in {\cal A}_B$.

In summary, we have adopted an operational definition of system, as a collection of observables together with states that suffice to provide all predictions pertaining to these observables. We have thereby identified the material system $A'$ and its bath, the transverse electromagnetic system $B$. 
The mathematical decomposition $({\cal A}={\cal A}_A\otimes{\cal A}_B,{\cal H}={\cal H}_A\otimes {\cal H}_B)$ is not defined directly in terms of observables and states but in terms {\em operators} and {\em vectors} whose physical meaning is different in each different frame of ${\cal H}$. In total we have defined four physically distinct subsystems $A$, $A'$, $B$, and $B'$ using two Hilbert space frames $X$ and $Y$. Only the standard quantum formalism has been employed and all four subsystems are equally well defined. With respect to frame $X$, the mathematical decomposition $({\cal A}_A\otimes{\cal A}_B,{\cal H}_A\otimes {\cal H}_B)$ yields the subsystems pair $(A,B)$ whereas with respect to frame $Y$ the same mathematical decomposition yields subsystem pair $(A',B')$. We refer to the pairs $(A,B)$ and $(A',B')$ as conjugate, because each requires the use of only a single Hilbert space frame. The physical dipole $A'$ and bath $B$, meanwhile, comprise the pair $(A',B)$, which we refer to as {\em non-conjugate} because it requires the use of two distinct frames $X$ and $Y$.

Energetic and entropic quantities can be defined for each quantum subsystem, $A, A', B$ and $B'$. We show in Sec.~\ref{therm} that physical laws can be provided to relate these quantities. The framework can therefore be understood as a generalisation in which the restriction of relating subsystem quantities defined using only a single Hilbert space frame is removed. 
We note finally that the approximation of the dipole-field example above to the case of a two-level dipole and a single mode is presented in appendix \ref{qrm}. We also note that it is straightforward to extend the example to multiple dipoles, multiple multi-mode baths, and to include arbitrary external potentials.

\subsection{Significance}\label{sig}

\begin{figure}[t]
\begin{minipage}{\columnwidth}
\begin{center}
\vspace*{-10mm}
\includegraphics[scale=0.24]{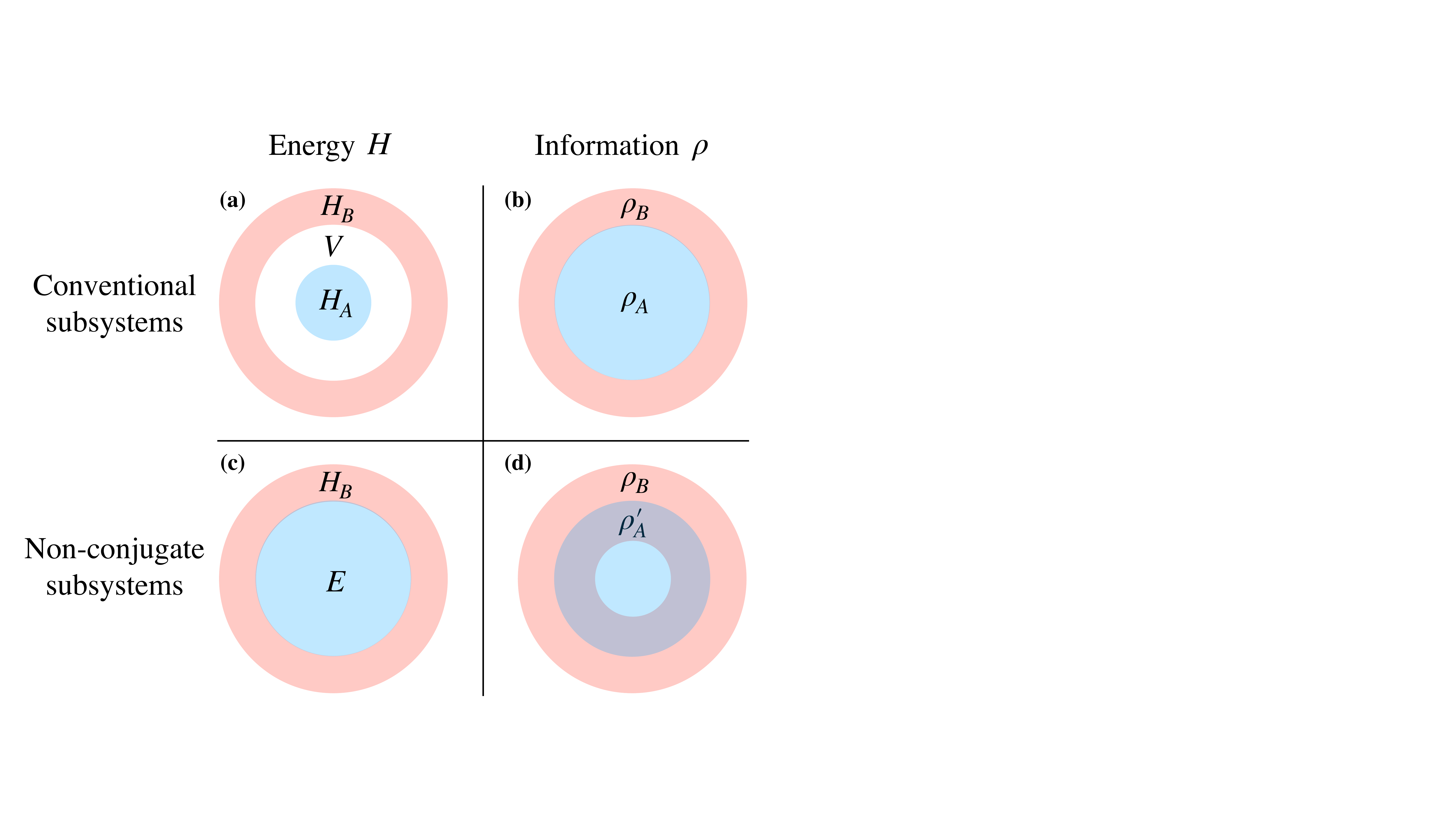}
\vspace*{-19mm}\caption{Conventional subsystems are complementary with respect to information but not energy. The information contained in the system $A$ conjugate to $B$ is defined as the information not contained in the bath $B$ as $\rho_A ={\rm tr}_B \rho$ (\textbf{b}), but as a result, the energy contained in the system $A$ cannot be the energy not contained in the bath; $H-H_B \not\in {\cal A}_A$ unless $V=0$ (\textbf{a}). In contrast, it is possible for non-conjugate subsystems $A'$ and $B$ to be complementary with respect to energy; $E_{A'}=H-H_B=H-E_B$ (\textbf{c}), but the information contained in the physical system $A'$ is not the information not contained in the physical bath $B$; $\rho_{A'} \neq {\rm tr}_B\rho$ (\textbf{d}).}\label{non_con}
\vspace*{-2mm}
\end{center}
\end{minipage}
\end{figure}

A non-conjugate decomposition provides different physical definitions of subsystems, resulting in different physical predictions and laws relating subsystem properties. For example, within the existing paradigm of open quantum systems theory, tracing out the bath $B$ is supposed to yield a description of the measurable open quantum system, but this supposition is false if the measurable open quantum system is $A'\neq A$. Complete statistical information pertaining to the physical system $A'$ is not simply obtained by tracing over the physical bath $B$, that is, $\rho_{A'} \neq {\rm tr}_B\rho$ (Fig~\ref{non_con}). One must instead trace over the conjugate subsystem $B'$ as $\rho_{A'}(t) = {\rm tr}_B\rho'$ where $\rho'(t)=R\rho(t)R^\dagger$. 

Let us suppose that one is then interested in predicting measurement signals pertaining to $A'$, such as its energy change. In 
the dipole-field example of Sec.~\ref{noncon}, solving the Heisenberg equation for ${\dot a}_{{\bf k}\lambda}(t)$ yields a solution in the form $a_{{\bf k}\lambda}(t) = e^{-i\omega_k t}a_{{\bf k}\lambda}(0) + a_{{\bf k}\lambda,s}(t)$ where $a_{{\bf k}\lambda,s}(t) = f_{A'}(t)$ and the operator $f_{A'}$ pertains to $A'$. This allows one to compute  averages such as $\langle a^\dagger_{{\bf k}\lambda,s}(t)a_{{\bf k}\lambda,s}(t)\rangle = {\rm tr}[\rho_{A'}(t) f_{A'}(t)^\dagger f_{A'}(t)]$ using one's reduced state $\rho_{A'}(t)$. But these averages are {\em not} correct predictions pertaining to $B$.

Since $H=E_{A'}+E_B$ and ${\dot H}=0$, we have;
\begin{align}\label{enab}
\Delta E_B=-\Delta E_{A'}.
\end{align}
This result can also be obtained by applying the divergence theorem to Poynting's theorem for local energy conservation. Specifically, the rates of change of the subsystem energies are given in terms of corresponding energy densities by ${\dot E}_{S} = \int d^3 x\, {\dot {\cal E}}_{S}$, $S=A',B$ where ${\dot {\cal E}}_{A'} = ({\bf J}_{\rm T}\cdot {\bf E}_{\rm T}+{\bf E}_{\rm T}\cdot {\bf J}_{\rm T})/2$, and ${\cal E}_B = ({\bf E}_{\rm T}^2+{\bf B}^2)/2$. Poynting's theorem states that ${\dot {\cal E}}_{A'} + {\dot {\cal E}}_B = -\nabla \cdot {\bf S}$ where ${\bf S}=({\bf E}_{\rm T}\times {\bf B}-{\bf B}\times {\bf E}_{\rm T})/2$. 
With respect to frame $Y$ the observables ${\cal E}_B,\,E_B$, and ${\bf S}$ do not belong in ${\cal A}_B$ and so they cannot be expressed as functions of the photonic operators $a_{{\bf k}\lambda}$ and $a^\dagger_{{\bf k}\lambda}$ alone. 

In Eq.~(\ref{enab}) one can understand the bath $B$ to be defined as the subsystem that, in the absence of external driving, is both necessary and sufficient to account for all energy changes in $A'$. Indeed, Eq.~(\ref{enab}) implies that
\begin{align}\label{f0}
Q&:={\rm tr}[(\rho_B(0)-\rho_B(t))E_B] \nonumber \\&= {\rm tr}[(\rho_{A'}(t)-\rho_{A'}(0))E_{A'}]=:U,
\end{align}
which is an instance of the first law of thermodynamics relating the non-conjugate subsystems $A'$ and $B$. Note that it is not possible to express the same law in terms of conjugate subsystem density operators and energy observables. If one is interested in determining the energy lost or gained by the measurable subsystem $A'$ then neither of the conjugate decompositions $(A,B)$ or $(A',B')$ will yield a strictly correct result. The former yields an incorrect reduced description of $A'$, while the latter yields an incorrect description of its energy exchanges with $B$. 
The non-conjugate subsystems $A'$ and $B$ are complementary with respect to energy, but it follows that they are not complementary with respect to information. Essentially the reverse is true for the conjugate subsystems $A$ and $B$. These underlying differences are depicted in Fig.~\ref{non_con}.

\subsection{Further Examples}\label{IV}

We provide further examples of physical models in which non-conjugate subsystems arise.\\

\subsubsection{Quantum Brownian motion}\label{bm}

The Caldeira-Leggett model of quantum Brownian motion describes a bound particle with bare Hamiltonian $H_A =  p_A^2/(2m)+\theta (r_A)$ interacting with a bath of quantum harmonic oscillators; $H_B = \sum_n \left(p_n^2/(2m_n)+m_n \omega_n^2 r_n^2/2\right)$. The Hamiltonian is given by \cite{caldeira_influence_1985,breuer_theory_2007}
\begin{align}\label{CL}
H_{\rm CL} &= {p_A^2\over 2m}+\theta(r_A) + \sum_n \left[{p_n^2\over 2m_n} + {m_n\omega_n^2 \over 2}\left(r_n-\kappa_n r_A \right)^2\right],\\
V_{\rm CL} &=  -\sum_n m_n\omega_n^2\kappa_n \left(r_A\otimes r_n -{\kappa_n \over 2}r_A^2\right) \nonumber \\ &= H_{\rm CL}-H_A-H_B
\end{align}
where the $\kappa_n$ are coupling parameters.

The system and bath particles share the interaction potential energy $V_{\rm CL}$ in the sense that $m{\ddot r}_A=-\partial_A (\theta + V_{\rm CL})$ and $m_n {\ddot r}_n = -m_n\omega_n^2 r_n - \partial_n V_{\rm CL}$. However, the position and momentum quadratures for each quantum oscillator can be relabelled by a local (unitary) Fourier transformation $S_n$ defined by
\begin{align}
S_n \psi(r_n) :=  \sqrt{{m_n\omega_n \over 2\pi}} \int_{-\infty}^\infty dr_n' \,e^{im_n\omega_n r_nr_n'}\psi(r_n'),
\end{align}
such that $S_n r_n S_n^\dagger = p_n/(m_n\omega_n)$ and $S_n p_n S_n^\dagger = -m_n\omega_n r_n$. This leaves the oscillator Hamiltonian $p_n^2/(2m_n)+m_n \omega_n^2 r_n^2/2$ unchanged, but it interchanges the kinetic and potential terms, such that $S_n (m_n \omega_n^2 r_n^2/2 )S_n^\dagger$ may be interpreted as kinetic energy. Letting $S_B = I_A\bigotimes_n S_n$, the Caldeira-Leggett Hamiltonian becomes $H'= S_BH_{\rm CL}S_B^\dagger$ and is given by
\begin{align}
H' =& {p_A^2\over 2m}+\theta(r_A) \nonumber \\ &+ \sum_n \left[{1\over 2m_n}\left(p_n-m_n\omega_n \kappa_n r_A\right)^2 + {m_n\omega_n^2 \over 2}r_n^2\right],\\
V' =& - \sum_n \omega_n \kappa_n \left(r_A\otimes p_n -{m_n\omega_n \kappa_n \over 2}r_A^2\right) \nonumber \\ =& \, H'-H_A-H_B
\end{align}
in which each bath oscillator is minimally-coupled to the system. Via the equation of motion $m_n {\dot r}_n = -im_n[r_n,H']=p_n -m_n\omega_n \kappa_n r_A$, the Hamiltonian $H'$ can now be written as the sum of kinetic and potential energies for each particle;
\begin{align}
H'=&E_{A'}+E_B,\\
E_{A'} =& {1\over 2}m{\dot r}_A^2 +\theta(r_A), \\
E_B =& \sum_n {m_n\over 2}\left[{\dot r}_n^2+\omega_n^2r_n^2\right]=H_B+V'.
\end{align}
The energy of the bath is defined as the sum of the total kinetic energy and total potential energy of the bath oscillators. This constitutes an expression of the theory in frame $Y$ whereby physical system $A'$ has observables represented by operators of the form $O_A\otimes I_B$. In particular $E_{A'}= H_A\otimes I_B$ is the energy of the physical system oscillator defined as the sum of its kinetic and potential energies. The state of the system oscillator at time $t$ is represented by $\rho_{A'}(t)={\rm tr}_B\rho'(t)$ where $\rho'(t)$ is the state of the composite at time $t$ in frame $Y$. The canonical momenta $p_n= m{\dot r}_n +m_n\omega_n \kappa_n r_A$ of the conjugate subsystem $B'$ possess no straightforward physical interpretation and are not equal to the mechanical momenta $m_n{\dot r}_n$ of the physical bath oscillators. Frame $X$, in which the physical bath oscillators comprise a subsystem $B$ is obtained using the unitary transformation
\begin{align}\label{RCL}
R = \exp\left[i\sum_n m_n\omega_n \kappa_n r_A \otimes r_n \right].
\end{align}
In particular, in frame $X$ we have $p_n=m_n{\dot r}_n$ and the bath energy is represented by $E_B =R^\dagger ( H_B+V')R = I_A\otimes H_B$. The reduced state of the physical bath is given by $\rho_B(t) ={\rm tr}_A\rho(t)$ where $\rho'(t)=R \rho(t) R^\dagger$. We see that in this model, only a non-conjugate decomposition $(A',B)$ defines {\em all} oscillators in a physically consistent way using their position and kinetic momentum observables. Unlike the non-conjugate decomposition $(A',B)$ which applies the same mechanical definition of ``oscillator" to both system and bath, each of the conjugate pairs $(A,B)$ and $(A',B')$ applies two different definitions of ``oscillator" in defining what is meant by ``system oscillator" versus what is meant by ``bath oscillator".

\subsubsection{Independent-boson and spin-boson models}

The independent-boson model consists of a TLS with raising and lowering operators $\sigma^\pm$, coupled to a bosonic reservoir. The Hamiltonian in frame $Y$ is \cite{nazir_modelling_2016}
\begin{align}\label{sb0}
H' =& \omega_m\sigma^+\sigma^- + \sigma^+\sigma^-\sum_{\bf k} {g_{\bf k}\over \sqrt{2}}(b^\dagger_{\bf k}+b_{\bf k}) \nonumber \\ &+ \sum_{\bf k} \omega_{\bf k} \left(b_{\bf k}^\dagger b_{\bf k}+{1\over 2}\right).
\end{align}
The model can be used to describe, for example, electron-phonon interactions in a quantum dot \cite{nazir_modelling_2016}. The phonon bath results from the dot substrate, which is coupled to excitations of a confined electron within the dot treated as a TLS with energy separation $\omega_m$. The polaron transformation
\begin{align}\label{polaron}
R = \exp\left[i\sigma^+\sigma^-\sum_{\bf k} {g_{\bf k}\over \omega} y_{\bf k}\right],
\end{align}
where $y_{\bf k}=i(b_{\bf k}^\dagger -b_{\bf k})/\sqrt{2}$, diagonalises $H'$. We label the polaron frame by $X$. The subsystems $A$, $A'$, $B$ and $B'$ can now be identified. To determine their physical meanings we write the Hamiltonian in Eq.~(\ref{sb0}) as
\begin{align}\label{sb1}
&H'=E_{A'}+E_B,\\
&E_{A'} = {\tilde \omega}_m \sigma^+\sigma^- =:H_A,\\
&E_B = \sum_{\bf k} {\omega_{\bf k}\over 2}\left( y_{\bf k}^2+ \left[x_{\bf k}+{g_{\bf k}\over \omega_{\bf k}}\sigma^+\sigma^-\right]^2\right),\label{EB}
\end{align}
where $x_{\bf k}=(b_{\bf k}^\dagger +b_{\bf k})/\sqrt{2}$ and ${\tilde \omega}_m =\omega_m-\sum_{\bf k} g^2_{\bf k}/(2\omega_{\bf k})$. If we adopt an interpretation in which physically measurable quantities are renormalised rather than bare, then when the confined electron is interacting with the phonon substrate, the measurable energy observable associated with the physical electronic TLS $A'$ possesses the polaron-shifted frequency ${\tilde \omega}_m$ rather than the bare frequency $\omega_m$. The associated energy operator is $E_{A'}$. 

The conjugate subsystem $B'$ has energy represented by the operator $I_A\otimes H_B$ in frame $Y$ where $H_B:= \sum_{\bf k} \omega_{\bf k} \left(b_{\bf k}^\dagger b_{\bf k}+{1\over 2}\right)$. On the other hand, the polaron transformation in Eq.~(\ref{polaron}) yields $R^\dagger(H' -H_A)R = H_B$ implying that the non-conjugate subsystem $B$ has energy $E_B$ represented by $I_A\otimes H_B$ in frame $X$. The bath $B$ within the polaron frame $X$ is dressed by the electronic TLS. It implicitly includes a deformation of the bare bath that occurs when the TLS is excited. Since $[R,H_A]=0$, the Hamiltonian in frame $X$ is $H = H_A+H_B$ such that the conjugate subsystems $A$ and $B$ are uncoupled.

The transformation to frame $X$ is less trivial if we consider the case of a driven quantum dot. This situation can be described using the spin-boson Hamiltonian obtained by adding to the independent-boson Hamiltonian in Eq.~(\ref{sb0}) a term $\Omega\sigma^x/2$ where $\Omega$ is the (semi-classical) laser driving strength. This gives \cite{nazir_modelling_2016}
\begin{align}
&H'=E_{A'}+E_B,\\
&E_{A'} = {\tilde \omega}_m \sigma^+\sigma^- + {\Omega\over 2}\sigma^x=:H_A,
\end{align}
with $E_B$ defined as in Eq.~(\ref{EB}). The Hamiltonian defined in this way is an expression in frame $Y$, which is rotating at the laser frequency and in which rapidly oscillating driving terms have been dropped. Applying the polaron transformation again yields $R^\dagger (H-H_A)R=H_B$, but now since $[R,\sigma^x]\neq 0$, we have $R^\dagger H_AR\neq H_A$. Instead, we obtain as the representation of the energy $E_{A'}$ in frame $X$
\begin{align}
R^\dagger H_A R = {\tilde \omega}_m\sigma^+\sigma^- + {\Omega\over 2}\left(D\sigma^++D^\dagger\sigma^-\right)
\end{align}
where $D:=\exp\left[\sum_{\bf k}{g_{\bf k}\over \omega_{\bf k}}(b_{\bf k}-b_{\bf k}^\dagger)\right]$. As in the case of the quantum Rabi Hamiltonian for a TLS, which is discussed in Appendix \ref{qrm}, this expression clearly possesses highly non-trivial dependence on the bosonic mode operators.


\section{Thermodynamics of non-conjugate subsystems}\label{therm}

Having provided motivating examples, we now discuss energies and entropies of non-conjugate subsystems in general. We prove first and second laws of thermodynamics for non-conjugate subsystems. We then derive non-trivial thermodynamic differences between conjugate and non-conjugate subsystems, and end with a simple illustrative example.

\subsection{Information}\label{cormi}



Given three quantum systems $A$, $B$, and $C$, the difference between the information available via measurements of $C$ and the combined information available via measurements of $A$ and $B$, is $I(A,B,C)=S(\rho_A)+S(\rho_B)-S(\rho_C)$ where $S(\rho):=-{\rm tr}(\rho\ln\rho)$. If $C$ is a composite system comprised of conjugate subsystems $A$ and $B$, then the information contained in $A$ ($B$) is the information not contained in $B$ ($A$), as expressed by the definitions ${\rm tr}_B\rho =\rho_A$ and ${\rm tr}_A\rho =\rho_B$ where $\rho:=\rho_C$ represents the state of $C$. In this case, $I(A,B,C)=I(\rho_A,\rho_B)$ where
\begin{align}
I(\rho_A,\rho_B) = S(\rho\|\rho_A\otimes \rho_B) \geq 0
\end{align}
in which $S(\rho\|\sigma):={\rm tr}(\rho\ln\rho)-{\rm tr}(\rho\ln\sigma)$ \cite{nielsen_quantum_2000}. For an initially uncorrelated state $\rho=\rho_A \otimes \rho_B$ the mutual information at a later time $t$ is
\begin{align}\label{mic}
0\leq I(\rho_A(t),\rho_B(t)) = \Delta S_A(t)+\Delta S_B(t),
\end{align}
which is an important identity for proving thermodynamic relations involving subsystem entropy changes, as will be seen in Sec.~\ref{therm}. The non-negativity of $I(\rho_A,\rho_B)$ tells us that the information associated with $C$ includes that associated with $A$ and $B$ as well as that associated with their correlations. If the subsystems are statistically independent then $I=0$ and if they are correlated then $I>0$. Viewed differently, learning about $A$ can only ever {\em reveal} information about conjugate subsystem $B$. This is a consequence of the trivial compatibility of subsystem observables specified within condition i).

In contrast, there is generally an overlap in the information associated with non-conjugate subsystems $A'$ and $B$. It is therefore possible for the total information associated with such non-conjugate subsystems to be larger than that associated with the composite. Viewed differently, this can be interpreted as demonstrating the capacity of one subsystem to {\em erase} information about the other. A measurement of $A'$ can immediately disturb non-conjugate subsystem $B$. Indeed, in Sec.~\ref{II} we identified this property as part of what might reasonably be taken to define interactions [condition ii)].

For non-conjugate subsystems $(A',B)$ the quantity most closely analogous to the mutual information of conjugate subsystems is $I(A',B,C)=I(\rho_{A'}.\rho_B)$ where
\begin{align}\label{minc}
I(\rho_{A'},\rho_B)&= S(\rho_{A'})+S(\rho_B)-S(\rho), \nonumber \\
&=S(\rho_{A'})+S(\rho_B)-S(\rho'),
\end{align}
This quantity is not, however, a non-negative relative entropy. It is bounded instead by $I(\rho_{A'},\rho_B)= S(\rho_{A'})-S(\rho_A)+I(\rho_A,\rho_B) \geq S(\rho_{A'})-S(\rho_A)$ or similarly by $I(\rho_{A'},\rho_B)= S(\rho_B)-S(\rho_B')+I(\rho_{A'},\rho_{B'})\geq S(\rho_B)-S(\rho_{B'})$. It follows that, even if $V(0)=0$ such that initially $A'=A$ and even if the initial state is uncorrelated, $\rho=\rho_A\otimes \rho_B$, the sum of non-conjugate subsystem entropy changes is not generally non-negative;
\begin{align}\label{mic2}
0\not \leq I(\rho_{A'}(t),\rho_B(t)) = \Delta S_{A'}(t)+\Delta S_B(t),
\end{align}
contrasting Eq.~(\ref{mic}). This will be seen in Sec.~\ref{further} to imply non-trivial differences between the thermodynamics of conjugate and non-conjugate subsystems.

Assuming $\rho=\rho_A\otimes \rho_B$ then $\rho_{A'} = \Phi(\rho_A) = \sum_{\lambda_B, \mu_B} K_{\mu_B\lambda_B} \rho_A K_{\mu_B\lambda_B}^\dagger$ where $\Phi$ is a quantum operation in which $K_{\mu_B\lambda_B} = \sqrt{\lambda_B}\bra{\mu_B}R\ket{\lambda_B}$ and where $\rho_B = \sum_{\lambda_B} \lambda_B\ket{\lambda_B}\bra{\lambda_B}$. The quantity $I(\rho_{A'},\rho_B)=S(\rho_{A'})-S(\rho_A)$ is negative if the operation $\Phi$ is entropy reducing. Such $\Phi$ certainly exist, an example is given in Appendix~\ref{JC}. Conversely, to a given a quantum operation $\Phi$ acting on a system $A$ it is possible to associate non-conjugate subsystems $A'$ and $B$, because any such operation may be written $\Phi(\rho_A) = {\rm tr}_B(R\rho_A\otimes \ket{\psi_B}\bra{\psi_B}R^\dagger)$ where $R$ is unitary \cite{nielsen_quantum_2000}.


To show in a general way how a measurement of $A'$ can increase the uncertainty in the state of system $B$, let us suppose that in frame $X$ the state of the composite is represented by $\rho = \rho_A \otimes \ket{\psi_B}\bra{\psi_B}$, such that the state of $B$ is pure. In frame $Y$ the state is entangled; $\rho' = R\rho_A \otimes \ket{\psi_B}\bra{\psi_B}R^\dagger$ and the system's energy is represented by an operator that we denote $H'_A\otimes I_B$. Let us assume for simplicity that $H_A'$ has non-degenerate spectrum; $H_A'\ket{\epsilon_i}=\epsilon_i\ket{\epsilon_i}$. An ideal projective measurement of the system $A'$'s energy yields the outcome $\epsilon_i$ with probability $p_i' = {\rm tr}(P_i\otimes I_B\rho')={\rm tr}(P_i\rho_{A'})$ where $P_i=\ket{\epsilon_i}\bra{\epsilon_i}$ and $\rho_{A'}={\rm tr}_B\rho'$. The subensemble for which outcome $\epsilon_i$ is found is represented in frame $Y$ by
\begin{align}
\sigma'_i = {1\over p_i'}P_i\otimes I_B\rho' P_i\otimes I_B = P_i \otimes \sigma_{iB'}
\end{align}
where $\sigma_{iB'}={\rm tr}_A\sigma'_i$. In frame $X$ this state is represented by $\sigma_i=R^\dagger  \sigma_i' R$. The final state of $A'$ is represented by $\sigma_{iA'}=P_i$. This state is pure (maximally informative), but since $\sigma_i$ is entangled the physical bath state represented by $\sigma_{iB} = {\rm tr}_A\sigma_i$ is mixed. 
A gain in information (reduction in uncertainty) about $A'$ by
\begin{align}
S(P_i)-S(\rho_{A'})= -S(\rho_{A'}) \leq 0,
\end{align}
has resulted in an increase in uncertainty about $B$ by
\begin{align}
S(\sigma_{iB})-S(\ket{\psi_B}\bra{\psi_B})= S(\sigma_{iB}) \geq 0.
\end{align}
For conjugate subsystems this is impossible. Following a measurement of $A$ 
the state of $B$ continues to be represented by the vector $\ket{\psi_B}$. 

For a non-selective measurement of system $A'$'s energy, the final state of the composite is represented in frame $Y$ by $\sigma' = \sum_i p'_i \sigma'_i$ and in frame $X$ by $\sigma = \sum_i p_i' \sigma_i$. The final state of the physical system $A'$ is represented by $\sigma_{A'} = \sum_i p_i'P_i$ and its entropy change is
\begin{align}
S(\sigma_{A'})-S(\rho_{A'}) = S(\rho_{A'}\|\sigma_{A'})\geq 0.
\end{align}
Although subsystems $A'$ and $B'$ are correlated, the measurement is non-selective, so the state of subsystem $B'$ remains unchanged; $\sigma_{B'} = {\rm tr}_A\sigma' = {\rm tr}_A \rho' = \rho_{B'}$. On the other hand, the final state of $B$ is represented by $\sigma_B =  \sum_i p_i' \sigma_{iB}$, incurring an increase in entropy 
\begin{align}
S(\sigma_B)-S(\ket{\psi_B}\bra{\psi_B}) = S(\sigma_B)\geq 0.
\end{align}
Such a change is impossible following a non-selective measurement of the conjugate system $A$. A simple example in which these general results can be illustrated is the Jaynes-Cummings model of quantum optics, which is discussed briefly in Appendix \ref{JC}.

\subsection{First and second laws}

Let us now consider a working system, labelled $A'$, interacting with a single thermal bath, labelled $B$. In a Hilbert space frame that we label $X$, we suppose that the total Hamiltonian is $H=H_A(t)+V(t)+H_B$ where $H_A(t)\in {\cal A}_A$ and $H_B\in {\cal A}_B$ are bare Hamiltonians and $V(t)$ is an interaction Hamiltonian. By assumption, the working system is accessible to measurement and control protocols whereas information about the bath is typically limited. We consider formulations of thermodynamics in which the energy of the working system is defined as $E_{A'}(t):=H_A(t)+V(t)$, such that $H(t)=E_{A'}(t)+E_B$ where $E_B:=H_B$ is the energy of the physical bath \cite{esposito_entropy_2010,esposito_second_2011,deffner_information_2013,strasberg_quantum_2017,strasberg_first_2021}. With these definitions the first law of thermodynamics becomes essentially automatic \cite{esposito_entropy_2010,esposito_second_2011,deffner_information_2013,strasberg_quantum_2017,strasberg_first_2021}. The second law is discussed and proven subsequently. 

\subsubsection{First law}\label{1law}

The first law of thermodynamics for the internal energy change of the working system $A'$ reads
\begin{align}\label{1st}
U = W+Q.
\end{align}
The total work $W$ is typically defined as the change in total energy; $W={\rm tr}[H(t)\rho(t)-  H(0)\rho(0)]$ such that ${d W/dt} =  {\rm tr}[\{{\dot H}_A(t)+{\dot V}(t)\}\rho(t)]$ \cite{esposito_entropy_2010,esposito_second_2011,deffner_information_2013,strasberg_quantum_2017,strasberg_first_2021}. The heat $Q$ is defined as minus the average change in energy of the bath, which is the part of the composite that is not directly controlled; $Q=-{\rm tr}[H_B\rho_B(t)-H_B\rho_B(0)] =-\Delta \langle E_B\rangle_t$ and ${d Q/dt} = -{\rm tr}[H_B{\dot \rho}(t)]$ \cite{esposito_entropy_2010,esposito_second_2011,deffner_information_2013,strasberg_quantum_2017,strasberg_first_2021,reeb_improved_2014,goold_nonequilibrium_2015}. 
It then follows from Eq.~(\ref{1st}) that
\begin{align}
U &= {\rm tr}[\{H_A(t)+V(t)\}\rho(t)-\{H_A(0)+V(0)\}\rho(0)]\nonumber  \\ &= \langle E_{A'}(t)\rangle -\langle E_{A'}(0)\rangle.
\end{align}
This confirms our definition of the working system's energy operator as $E_{A'}(t) = H_A(t)+V(t)$, which is also the definition arrived at in a number of works within the thermodynamics literature \cite{esposito_entropy_2010,esposito_second_2011,deffner_information_2013,strasberg_quantum_2017,strasberg_first_2021}. Note however, that here we have started with the definition $E_{A'}(t)=H_A(t)+V(t)$, which does not fix $Q$ and $W$ beyond the requirement that Eq.~(\ref{1st}) is satisfied. The definitions of $Q$ and $W$ above are particular examples, but they are not the only consistent definitions.


\subsubsection{Temperature and equilibrium}

The second law of thermodynamics in its most basic and fundamental form states that in any physical process the thermodynamic entropy $S_{\rm th}$ (yet to be defined here) of a closed system does not decrease;
\begin{align}\label{2gen}
\Delta S_{\rm th} \geq 0
\end{align}
such that a closed system tends to a state of maximum entropy (equilibrium). 
A closed system is one that has no explicit interactions with anything else. Controlled processes like measurements and driving may  be considered as external, i.e., without the devices and associated interactions being treated explicitly. Typically, the closed system is assumed to consist of a working system $A'$, which is measurable, and at least one thermal bath. Here we are considering the case of a single thermal bath $B$. 
The definition of entropy $S_{\rm th}$ should relate in some way to the information available to an experimenter who performs measurements. Once $\Delta S_{\rm th}$ can be connected to physically measurable properties the second law becomes a meaningful and useful physical statement.

A suitable such definition of $S_{\rm th}$ is given in Sec.~\ref{2p}. We note first that in order to relate entropy and energy a quantity with dimensions inverse energy is required and the one employed almost universally is inverse temperature $\beta$. Temperature can be derived by relating information and energy via the notion of equilibrium \cite{jaynes_information_1957}. Since $\rho$ is sufficient to provide complete statistical information regarding any observable, the Von-Neumann entropy $S(\rho) = -{\rm tr}(\rho\ln\rho)$ can be interpreted as quantifying the information available to an observer who is able to perform any measurement. If the system possesses fixed average energy $\langle E \rangle ={\rm tr}(\rho E)$ and if one maximises $S(\rho)$ subject to this constraint, i.e., if one minimises the Lagrangian $F(\rho,\beta) =\langle E \rangle-\beta^{-1}S(\rho)$ where $\beta$ is a Lagrange multiplier, then one obtains $\rho = \rho^{\rm eq}(\beta) := e^{-\beta E}/Z$ where $Z= {\rm tr} e^{-\beta E}$. The state $\rho^{\rm eq}(\beta)$ is an equilibrium state of the system with temperature $\beta^{-1}$. The Lagrangian $F(\rho,\beta)$ is called the free energy, which by construction satisfies $F(\rho^{\rm eq}(\beta),\beta)\leq F(\rho,\beta)$ for all $\rho \in {\cal D}$. Note that the entropy $S(\rho)$ comes from information theory and should not tacitly be equated with thermodynamic entropy. The above calculation indicates only that these entropies should coincide at thermal equilibrium.

\subsubsection{Consistent identification of observables and states}\label{cons}


In conjunction with the definition $E_{A'}(t):=H_A(t)+V(t)$, previous works have assumed the standard definition of the reduced system state; $\rho_A(t):={\rm tr}_B\rho(t)$ where $\rho(t)$ represents the state of the composite in frame $X$ \cite{esposito_entropy_2010,esposito_second_2011,strasberg_quantum_2017,strasberg_first_2021}. However, the two definitions $E_{A'}(t)$ and $\rho_A(t)$ are incompatible, because the average of $E_{A'}(t)$ cannot be computed using $\rho_A(t)$ alone, and yet in any physical theory, the {\em state} of a system must by definition suffice to provide complete statistical information for all system observables. In other words, we have that $\rho_A\in {\cal D}_A$ whereas ${\cal A} \ni H_A(t)+V(t)  \not \in {\cal A}_A$ and yet unlike $({\cal D}_A,{\cal A}_A)$, the pair $({\cal D}_A,{\cal A})$ is not a quantum system. In particular, the Von Neumann entropy $S(\rho_A(t))$ defined in terms of $\rho_A(t)$ cannot be said to quantify the information available to an experimenter able to make measurements of the working system $A'$.

If $E_{A'}(t)$ and $E_B$ are assumed to be the energies of the physical working system $A'$ and its bath $B$ respectively, and if $A'$ and $B$ are also assumed to be quantum subsystems, 
then they must be non-conjugate. Specifically, there must exist a rotation $R(t)$ mapping from frame $X$ to a frame $Y$, such that $R(t)[H_A(t)+V(t)]R(t)^\dagger= H_A'(t)\otimes I_B$ where $H_A'(t)$ is an operator that belongs in ${\cal A}_A$ and which represents the energy observable $E_{A'}(t)$ in frame $Y$. Entropy, as a measure of information available to an experimenter who measures $A'$, is then correctly defined using $\rho_{A'}(t)$. As we will show, this fact implies non-trivial corrections to the basic thermodynamic quantities found within previous formulations \cite{esposito_entropy_2010,esposito_second_2011,deffner_information_2013,strasberg_quantum_2017,strasberg_first_2021}. 

\subsubsection{Second law}\label{2p}

For a closed system with unitary dynamics, the Von Neumann entropy is invariant, $\Delta S = 0$. It is not generally suitable as a definition of thermodynamic entropy, because it refers to the information available to an experimenter able to perform any measurement that is possible in principle rather than referring to the specific measurements that the experimenter does perform. A more suitable definition is provided by observational entropy, which is defined in terms of the measurements made. Here we briefly review the definition and apply it to prove the second law in its most basic form [inequality~(\ref{2gen})]. We refer the reader to Refs.~\cite{safranek_quantum_2019,safranek_quantum_2019-1,safranek_classical_2020,strasberg_first_2021} for further details. 

Consider an observable ${\cal O}$ represented by a Hermitian operator $O = \sum_i o_i\Pi_i$ such that ${\cal C}=\{\Pi_i\}$ defines a coarse-graining of ${\cal H}$, that is complete; $\sum_i \Pi_i =I$. The probability that outcome $o_i$ is obtained in a measurement of ${\cal O}$ is $p_i ={\rm tr}(\Pi_i\rho)$ where $\rho$ represents the (micro)state of the system. A measurement with outcome labelled by $i$ corresponds to a portion of ${\cal H}$ that represents states compatible with this measurement outcome. Thus, $i$ labels a macrostate of the system with state-space ``volume" $V_i = {\rm tr} \Pi_i$. To a given microstate compatible with the macrostate $i$, Boltzmann associated an entropy equal to the log of the volume of the macrostate $i$. One therefore defines the Boltzmann entropy as ${\cal S}_i = \ln V_i$. The {\em observational entropy} is defined by
\begin{align}\label{obs}
S_{\cal C}(\rho) = -\sum_i p_i \ln {p_i\over V_i} = S_{\rm Sh}\{p_i\}+\sum_i p_i {\cal S}_i
\end{align}
where $S_{\rm Sh}\{p_i\} :=-\sum_i p_i \ln p_i$ is the Shannon entropy of the distribution $\{p_i\}$. If $\rho=\sum_i p_i\Pi_i$ then this term coincides with the Von Neumann entropy $S(\rho)$. The first term in Eq.~(\ref{obs}) therefore represents the expected information regarding the macrostate of the system obtained from a measurement with coarse-graining ${\cal C} =\{\Pi_i\}$. The second term represents the remaining average uncertainty regarding the state of the system once the measurement outcome is known \cite{safranek_quantum_2019-1}.

To describe a sequence of measurements a sequence of coarse-grainings ${\cal C}_1,...,{\cal C}_n$ must be specified. This defines a multimacrostate ${\bf i}=(i_1,\dots,i_n)$ with volume $V_{i_1...i_n}={\rm tr}(\Pi_{i_1}\dots\Pi_{i_n}\dots \Pi_{i_1})$. The quantity $p_{i_1...i_n} = {\rm tr}(\Pi_{i_n}\dots\Pi_{i_1}\rho \Pi_{i_1}\dots \Pi_{i_n})$ is the joint probability that an outcome labelled by $i_1$ is obtained in the first measurement, and outcome labelled by $i_2$ is obtained in the second measurement, and so on, up to the $n'th$ measurement. Note that we do not require the measurements to be compatible, that is, in general $[\Pi_{i_m},\Pi_{i_n}]\neq 0$. The observational entropy can in this case be defined by  \cite{safranek_quantum_2019-1}
\begin{align}\label{obs}
S_{{\cal C}_1...{\cal C}_n}(\rho) = -\sum_{\bf i} p_{i_1...i_n} \ln {p_{i_1...i_n}\over V_{i_1...i_n}}.
\end{align}
One expects intuitively that the least uncertainty possible is given by the Von Neumann entropy, which is indeed the case;
\begin{align}\label{obsineq}
S_{{\cal C}_1...{\cal C}_n}(\rho) \geq S(\rho).
\end{align}
The proof involves a straightforward application of Jensen's inequality to the concave function $f(x)=-x\ln x$. Details can be found in Ref.~\cite{safranek_quantum_2019-1}.

Let us now suppose that an experimenter is able to perform measurements corresponding to some coarse-graining ${\cal C}_{A'}$, on a working system labelled $A'$, as well as energy measurements corresponding to a coarse-graining ${\cal C}_{E_B}$, of the surrounding bath labelled $B$. This scenario was encountered in the dipole-field example of Sec.~\ref{noncon}. It was noted there that the relevant energy observables obey Poynting's theorem [Eq.~(\ref{enab})], which is an operator form of the first law of thermodynamics [Eq.~(\ref{f0})], and that they are {\em incompatible}. We must therefore also allow this to be the case when considering the second law. 
In particular, we allow the subsystems $A'$ and $B$ to be non-conjugate. However, following conventional open quantum systems theory \cite{breuer_theory_2007,de_vega_dynamics_2017}, we consider an initial state of the form $\rho= \rho_{A}(0)\otimes \rho_B^{\rm eq}(\beta)$ in which the bath is at thermal equilibrium with temperature $\beta^{-1}$. This is justified for arbitrary coupling strengths provided that at $t=0$ the system and bath are non-interacting and subsequently brought into contact at some time $t>0$. It follows that at $t=0$ the subsystems are conjugate with total energy $H(0)= E_{A'}(0)+E_B = H_A\otimes I_B+I_A\otimes H_B$. At later times $H(t)=E_{A'}(t)+E_B$ but in general $[E_{A'}(t),E_B]\neq0$.

Since the experimenter is able to make measurements ${\cal C}_{A'}$ we can suppose that $\rho_{A'}(0)=\rho_A(0)$ is of the form $\ket{s}\bra{s} \in {\cal C}_{A'}$, which represents a state in which the value of the measured observable is known. For simplicity we will assume that the spectrum $\{s\}$ of this observable is non-degenerate. We also suppose that the available bath energy measurements are of sufficiently high resolution that ${\cal S}(\beta) = S(\rho^{\rm eq}_B(\beta)) \approx S_{{\cal C}_{E_B}}(\rho^{\rm eq}_B(\beta))$ \cite{strasberg_first_2021}. This means that the initial bath observational entropy is indeed that of a canonical ensemble at equilibrium. At time $t$ the experimenter is able to perform measurements ${\cal C}_{A'}^t$ on the working system $A'$. The possible outcomes, $\{s_t\}$, depend on $t$ in general.

Since at $t=0$ we have $S_{C_{A'}\otimes C_{E_B}}(\rho) = S(\rho)$ and moreover $S(\rho)=S(\rho(t))$, it follows from inequality (\ref{obsineq}) that the change in observational entropy corresponding to the measurement sequence ${\cal C}_{A'}^t, {\cal C}_{E_B}$ satisfies
\begin{align}
\Delta S_{{\cal C}_{A'}{\cal C}_{E_B}} := S_{{\cal C}_{A'}^t{\cal C}_{E_B}}(\rho(t)) - S_{{\cal C}_{A'}{\cal C}_{E_B}}(\rho) \geq 0
\end{align}
in which we identify the left-hand-side as a thermodynamic entropy change. Similarly the thermodynamic entropy change corresponding to the measurement sequence ${\cal C}_{E_B},{\cal C}_{A'}^t$ is non-negative. This generalises the second law proved in Ref.~\cite{strasberg_first_2021}. 

The observational entropy is defined directly in terms of measurements performed, which makes it an immediately relevant thermodynamic entropy provided that a consistent identification of which observables the experimenter measures is made (cf. Sec.~\ref{cons}). For example, if the energy $E_{A'}(t)=H_A(t)+V(t)\not \in {\cal A}_A$ is measured and $[E_{A'}(t),E_B]\neq 0$, then the relevant observational entropy is evidently that associated with two incompatible observables. We remark before continuing that a non-negative thermodynamic entropy change as in inequality (\ref{2gen}) is sometimes referred to as entropy production. Here however, we reserve the latter term for a different quantity that will be considered in the following section.

\subsection{Further thermodynamic relations}\label{further}

The observational entropy provides a natural and operationally relevant second law. For system $A$ and conjugate bath $B$ some further relations can also be provided, which Ref.~\cite{strasberg_first_2021} refers to as a hierarchy of second laws. We briefly review these results, which will allow us to understand the key differences obtained when we subsequently consider the {\em measurable} working system $A'$ and non-conjugate bath $B$.

\subsubsection{Thermodynamic relations for the entropy of system $A$}\label{prev}

Since the system and bath are initially uncorrelated we have $S_{{\cal C}_{A}^0{\cal C}_{E_B}}(\rho(0)) = S_{{\cal C}_A^0}(\rho_A(0))+S_{{\cal C}_{E_B}}(\rho_B(0))$, using which one immediately obtains an additive law analogous to Eq.~(\ref{mic}) as $\Delta S_{{\cal C}_A^t} + \Delta S_{{\cal C}_{E_B}} = I_{{\cal C}_A^t{\cal C}_{E_B}}(\rho_A(t),\rho_B(t)) + \Delta S_{{\cal C}_{A}^t{\cal C}_{E_B}} \geq \Delta S_{{\cal C}_{A}^t{\cal C}_{E_B}} \geq 0$ where the classical mutual information
\begin{align}\label{miobs}
I_{{\cal C}_A^t{\cal C}_{E_B}}(\rho_A(t),\rho_B(t)) &:= \sum_{s_t ,\epsilon_B} p_{s_t \epsilon_B}(t) \ln \left[ p_{s_t \epsilon_B}(t) \over p_{s_t}(t)p_{\epsilon_B} (t) \right] \nonumber \\ &\geq 0
\end{align}
quantifies the correlations within the measurement result $(s_t,\epsilon_B)$.

For incompatible measurements ${\cal C}_A'$ and ${\cal C}_B$ the situation is different. In this case there is no obvious way to express the quantity $I_{{\cal C}_{A'}{\cal C}_{B}}(\rho_{A'},\rho_B) := S_{{\cal C}_{A'}}(\rho_{A'})+S_{{\cal C}_{B}}(\rho_{B})-S_{{\cal C}_{A'}{\cal C}_B}(\rho)$ as a non-negative Kullback-Leibler divergence. This is analogous to the case of $I(\rho_{A'},\rho_B)$ given in Eq.~(\ref{minc}). Since we have already discussed $I(\rho_{A'},\rho_B)$ in Sec.~\ref{cormi}, we omit a similar analysis of $I_{{\cal C}_{A'}{\cal C}_{B}}(\rho_{A'},\rho_B)$. We note however that even when dealing with a macroscopic working system (``sys") and heat bath (``$B$"), one expects to obtain an extensive entropy of the form $S_{\rm th} = S_{\rm th, sys}+S_{\rm th, B}$ only as an approximation, albeit a very good one in many contexts. Yet, only when such a relation holds does the fundamental second law $\Delta S_{\rm th} \geq 0$ become the additive law $\Delta S_{\rm th, sys}+\Delta S_{\rm th, B} \geq 0$ and so the latter inequality is not fundamental in general.

Defining an effective inverse temperature $\beta^*(t)$ by the equation ${\rm tr}[\rho_B(t)H_B] = {\rm tr}[\rho_B^{\rm eq}(\beta^*(t))H_B]$, it is straightforward to show that \cite{strasberg_first_2021}
\begin{align}\label{bobs}
\Delta S_{{\cal C}_{E_B}} \leq \int dU_B(s) \beta^*(s) = -  \int dQ(s) \beta^*(s) 
\end{align}
where $U_B(t) = \Delta \langle E_B\rangle_t = {\rm tr}[\rho_B(t)H_B] -{\rm tr}[\rho_B(0)H_B]$ is the change in internal energy of the bath and $Q(t)=-U_B(t)$ as in Sec.~\ref{1law}. If the bath temperature hardly changes from its initial equilibrium value $\beta$, i.e., $\beta^*(s)\approx \beta$ for all $s\in [0,t]$, then inequality~(\ref{claus}) becomes
\begin{align}\label{bobs2}
\Delta S_{{\cal C}_{E_B}} \leq -\beta Q.
\end{align}
For conjugate system and bath the additive law $\Delta S_{{\cal C}_A^t} + \Delta S_{{\cal C}_{E_B}}\geq 0$ and inequality (\ref{bobs}) imply that
\begin{align}\label{claus}
\Delta S_{{\cal C}^t_{A}} \geq \int dQ(s) \beta^*(s)
\end{align}
and for $\beta^*(s) \approx \beta$
\begin{align}\label{claus2}
\Delta S_{{\cal C}^t_{A}} \geq \beta Q.
\end{align}
Inequalities (\ref{claus}) and (\ref{claus2}) are nonequilibrium Clausius-type relations between the system's thermodynamic entropy change and the heat $Q$, which hold for conjugate system and bath.  

Rather than considering the observational entropy, Ref.~\cite{esposito_entropy_2010} provides elegant laws for Von Neumann entropy changes. It is assumed that the system energy $E_{A'}$, heat $Q$, and work $W$ are defined as in Sec.~\ref{1law}. 
The difference $\Sigma : = \Delta S_A - \beta Q$ between the Von Neumann entropy change of the subsystem $A\neq A'$ and the thermal entropy $\beta Q$, is then found to be \cite{esposito_entropy_2010} 
\begin{align}\label{entprod}
&\Sigma = S(\rho(t)\|\rho_A(t)\otimes \rho_B^{\rm eq}(\beta)) \geq 0,
\end{align}
which quantifies the distance between the true state $\rho(t)$ and the hypothetical uncorrelated state $\rho_A(t)\otimes \rho_B^{\rm eq}(\beta)$ in which the bath $B$ has remained at equilibrium.  For an initial state $\rho_A(0)=\ket{s}\bra{s}$ diagonal at $t=0$ in some given measurement basis ${\cal C}_A$ we have $\Delta S_{{\cal C}^t_A} \geq \Delta S_A = \Sigma + \beta Q \geq \beta Q$, which gives a tighter bound on $\beta Q$ than the microscopic Clausius inequality (\ref{claus2}). 

Eq.~(\ref{entprod}) is mathematically exact, having been derived without approximation. Naturally, $\Sigma$ is bounded from below by the non-negative mutual information, which equals the sum of the conjugate subsystem entropy changes for an initially uncorrelated state [Eq.~(\ref{mic})]. The difference $[\Sigma -I(\rho_A(t),\rho_B(t))]/\beta$ is the bath free energy change, that is, $\beta \Delta F_B= S(\rho_B(t)\| \rho^{\rm eq}(\beta)) \geq 0$ where $F_B :=\langle E_B\rangle -\beta^{-1} S(\rho_B)$. The inequality $\Delta F_B \geq 0$ is equivalent to $\Delta S_B(t) \leq -\beta Q$, which is a looser lower bound on $-\beta Q$ than inequality (\ref{bobs2}). 

An important corollary specifying the maximum extractible work is \cite{esposito_second_2011,strasberg_quantum_2017} 
\begin{align}\label{workb}
W = \beta^{-1}\Sigma + \Delta F_A \geq \Delta F_A,
\end{align}
where $F_A : = \langle E_{A'}\rangle - \beta^{-1} S(\rho_A)$. Note that $F_A$ is neither the free energy of $A$ nor $A'$, but rather a hybrid quantity that combines the average energy of $A'$ with the entropy of $A$. It does not possess the defining properties of free-energy. For example, if $\rho$ is such that the state of subsystem $A$ is thermal, $\rho_A^{\rm eq}(\beta):=e^{-\beta H_A}/Z_A$, then $F_A = {\rm tr}[V\rho] +\beta^{-1}\ln Z_A \neq \beta^{-1}\ln Z_A$, where the right-hand-side is the equilibrium free energy of $A$. However, by assuming that $V(t)=0$ the subsystems $A$ and $A'$ coincide at time $t$, such that $F_A(t)$ is then the free energy of $A=A'$, which is minimised at equilibrium. Under this assumption, using Eq.~(\ref{workb}) one obtains
\begin{align}\label{wirr}
W-\Delta F_A^{\rm eq} = \beta^{-1}(\Sigma + \Delta I_A) \geq \beta^{-1}\Delta I_A
\end{align}
where $\beta^{-1}I_A(t): = \beta^{-1}S(\rho_A(t)\|\rho_A^{\rm eq}(\beta, t)) = F_A(t)-F_A^{\rm eq}(t)$ can be interpreted as quantifying the information that must be processed in transforming from the equilibrium state, represented by $\rho_A^{\rm eq}(\beta,t)  = e^{-\beta H_A(t)}/Z_A(t)$, to the actual state at time $t$, represented by $\rho_A(t)$. The authors of Ref.~\cite{esposito_second_2011} interpret Eq.~(\ref{wirr}) as a nonequilibrium Landauer principle. It states that the difference between the change in total energy $W$ and the free energy change of corresponding initial and final equilibrium distributions of $A$, is bounded by the {\em information change} $\Delta I_A$ associated with $A$.\\


\subsubsection{Thermodynamic relations for the entropy and the free energy of the system $A'$}\label{meas}

Thermodynamic laws are useful only insofar as the quantities appearing therein are physically relevant. 
For example, if $\rho_A(t)$ were sufficient to provide statistical predictions for the measurable working system $A'$, then Eq.~(\ref{entprod}) would tell us that the decrease in information available to the experimenter is at least as large as $\beta$ times the heat gained from $B$, while  inequality (\ref{workb}) would tell us that the maximum extractible work is given by the decrease in free energy of the measurable system. The results of Sec.~\ref{prev} {\em do not} constitute these useful physical statements, because 
$\rho_A(t)$ does not suffice to provide the averages of the physically relevant observables that define the working system $A'$, such as the energy $E_{A'}$. 

In order to provide {\em useful} thermodynamic laws for the working system $A'$ we must consider  the entropy and free energy {\em of this system}, which are defined by
\begin{align}
&S(\rho_{A'}(t)) \neq  S(\rho_A(t)),\\
&F_{A'}:=\langle E_{A'} \rangle -\beta^{-1}S(\rho_{A'}) \neq F_A.\label{fa'}
\end{align}
Since for non-conjugate system and bath one does not generally have an additive law $\Delta S_{{\cal C}_A^t} + \Delta S_{{\cal C}_{E_B}}\geq 0$, one cannot expect inequalities of the form in (\ref{claus}) and (\ref{claus2}) to hold in general. Indeed, by considering $S_{A'}$ and $F_{A'}$ instead of $S(\rho_A)$ and $F_A$, we will obtain a novel correction, which can either be interpreted as a correction to the naively expected thermodynamic relations, or instead as a correction to the naively expected definitions of heat and work. 

We begin by defining the naive entropy production ${\tilde \Sigma}:=\Delta S_{A'}(t)-\beta Q$. 
A straightforward calculation yields
\begin{align}\label{entprod2}
{\tilde \Sigma}= -S(\rho(t))-{\rm tr}[\rho'(t)\ln \rho_{A'}(t)]-{\rm tr}[\rho(t)\ln \rho_B^{\rm eq}(\beta)].
\end{align}
If the Hilbert space frames $X$ and $Y$ are locally connected, $R(t)=R_A(t)\otimes R_B(t)$, then the subsystems are conjugate and we recover Eq.~(\ref{entprod}). Another sufficient condition in order that the right-hand-side of Eq.~(\ref{entprod2}) is non-negative is that $[R(t),H_B]=0$, in which case using $\rho(t)=R(t)^\dagger \rho'(t)R(t)$ we find that ${\tilde \Sigma}=\Sigma'$ where
\begin{align}\label{entprod3}
\Sigma' := S(\rho'(t)\|\rho_{A'}(t)\otimes \rho_B^{\rm eq}(\beta)).
\end{align}
In general ${\tilde \Sigma}$ cannot be written as a relative entropy as in Eq.~(\ref{entprod3}), but a straightforward calculation reveals that we 
can write $\Delta S_{A'}(t)$ as
\begin{align}\label{entprod4}
\Delta S_{A'}(t) = \Sigma' +\beta Q'
\end{align}
where
\begin{align}
&Q' := -{\rm tr}[H_B\rho'(t)-H_B\rho'_B(0)] \nonumber \\  & \hspace*{3.9mm}= -{\rm tr}[H_B\rho'(t)-H_B\rho_B^{\rm eq}(\beta)] =Q +\delta Q ,\label{Qp}\\
& \delta Q := {\rm tr}[\rho(t)\delta H_B(t)],\label{delhb} \\
&\delta H_B(t) := H_B-R(t)^\dagger H_BR(t).\label{delHB}
\end{align}
Furthermore, using Eqs.~(\ref{fa'}) and (\ref{entprod4}), we obtain analogously to Eq.~(\ref{workb})
\begin{align}\label{fen}
W'- \Delta F_{A'} = {\beta}^{-1}\Sigma' \geq 0,
\end{align}
where  $W' := W-\delta Q$.

Mathematically, Eq.~(\ref{entprod4}) is nothing but the previous result $\Delta S_{A}(t) = \Sigma +\beta Q$ with $\rho(t)$ replaced everywhere by $\rho'(t)$. Physically, this difference is important,  because in frame $Y$, the operator $H_B$ does not represent the energy $E_B$, but a different observable that we will denote ${\cal E}_B$. The energy $E_B$, meanwhile, is represented by the operator $R(t)H_BR^\dagger(t)$. The correction $\delta Q$ is the average difference between these observables at time $t$; $\delta Q  =Q'-Q= \langle E_B - {\cal E}_B\rangle_t$.  The operator $\delta H_B(t)$ in Eq.~(\ref{delhb}) is of the same order as $V(t)$, therefore $\beta \delta Q$ is expected to be small for weak-coupling, but it should become significant for sufficiently strong-coupling and low temperatures.

\subsection{Meaning of $\delta Q$ and its physical implications}
 
Assuming an initially uncorrelated state, one obtains the (essentially trivial) additive relation (\ref{mic}) for conjugate subsystem Von Neumann entropy changes, but Eq.~(\ref{mic2}) shows that such an inequality does not generally hold for the sum $\Delta S_{A'}+\Delta S_B$ even if $A'$ and non-conjugate bath $B$ are initially non-interacting and uncorrelated. This results in the correction $\delta Q$, which implies that either  \textbf{a}) one no longer has a microscopic Clausius-type inequality of the form $\Delta S_{\rm sys} \geq \beta Q$ and associated free energy bound of the form $W \geq \Delta F_{\rm sys}$, that is, the maximum extractable work is {\em not} the decrease in the measurable system's free energy, or else, \textbf{b}) the naive definitions of heat and work, $Q$ and $W$, fail, and must be replaced by $Q'$ and $W'$ respectively; the physical heat is no longer simply minus the change in bath energy and the physical work is no longer simply the average change in total energy.

Furthermore, from Eq.~(\ref{fen}) we obtain
\begin{align}\label{wirr2}
W'-\Delta F_{A'}^{\rm eq} = \beta^{-1}(\Sigma'+ \Delta I_{A'}) \geq \beta^{-1}\Delta I_{A'}
\end{align}
of which Eq.~(\ref{wirr}) is a special case obtained by assuming that $V(t)=0$, such that $A=A'$ at time $t$ (and so $F_{A'}=F_A$, $I_{A'}=I_A$ and $W'=W$). Thus, use of the genuine free energy $F_{A'}$ rather than the hybrid quantity $F_A$ allows one to prove a general Landauer-type inequality (\ref{wirr2}) for non-conjugate subsystems, that cannot be proven for conjugate subsystems except in the specific case that the two definitions coincide. This bound involves the modified work $W'$, such that  $W$ is instead bounded by $\beta^{-1}\Delta I_{A'}+\delta Q$. Therefore, again we see that either \textbf{a}) $\delta Q$ constitutes a non-trivial correction to the expected thermodynamic relation, that is, the difference $W-\Delta F_{A'}^{\rm eq}$ between the work and equilibrium free-energy change is not simply bounded by the information change $\Delta I_{A'}$ associated with the working system $A'$, or else, \textbf{b}) heat and work must be non-trivially redefined as $Q'=Q+\delta Q$ and $W'=W-\delta Q$.

The physical significance of these implications is especially evident when considering the work. The free-energy bounds (\ref{fen}) and (\ref{wirr2}) which hold for $W'$, but not for $W$ are physically important thermodynamic relations, and yet the definition of work as $W$ is usually assumed to be unambiguous, not least because a redefinition  leads to similarly unexpected thermodynamics. For example, by supposing that the work done is $W' = W-\delta Q$ then even if there is no net total energy change, $W=0$, an amount of work $\delta Q$ would be extractible. Thus, we have shown that not all of the naively expected thermodynamic relations can be simultaneously satisfied by a {\em consistent} framework in which the working system's energy is $E_{A'}(t):=H_A(t)+V(t)$. All of the naively expected relations are simultaneously satisfied within the framework of Sec.~\ref{prev}, only because that framework is internally {\em inconsistent}. Specifically, the definition of the working system's density operator as being $\rho_A(t)$ is incompatible with the definition of the working system's energy as being $E_{A'}(t)=H_A(t)+V(t)$ (see Sec.~\ref{cons}). 

Of course, identifying the correct definition of the working system's energy is a hotly debated topic in quantum thermodynamics. The only alternative to accepting at least one of the non-trivial physical implications \textbf{a} or \textbf{b} would be a universal rejection of the definition $E_{A'}(t):=H_A(t)+V(t)$, despite that it has been independently derived in a number of studies in quantum thermodynamics \cite{esposito_entropy_2010,esposito_second_2011,deffner_information_2013,strasberg_quantum_2017,strasberg_first_2021}. Although different definitions may be appropriate in different contexts, we believe that the present results progress the discussion. It was found in the example of Sec.~\ref{noncon} that the definition $E_{A'}:=H_A+V$ is implied by physical equations of motion. It was noted in Sec.~\ref{sig} that only this definition is the same one provided by Poynting's theorem, which expresses the local conservation of energy, as implied by Noether's theorem and time-translation invariance. In this example therefore, the definition appears to be essentially beyond dispute on both experimental and theoretical grounds, provided of course that one's premise for defining what is meant by ``working system" is {\em operational}, as it must obviously be. Moreover, beyond the example in Sec.~\ref{noncon}, we subsequently demonstrated that this definition is physically well-motivated in several other specific models.

The only additional non-trivial assumption we have made is that $V(0)=0$, which allows us to assume an initial thermal bath and thereby introduce the temperature $\beta^{-1}$. This is a common assumption, which does not restrict the coupling strength but does require that we can view the interaction as switchable. It is an important special case of the more general setting in which one allows completely arbitrary initial states and completely arbitrary $V(t)$. We leave for further work this extension, as well as the task of developing further thermodynamic relations.

\subsection{Example}\label{exa}

\begin{figure}[t]
\begin{minipage}{\columnwidth}
\begin{center}
\hspace*{-2mm}\includegraphics[scale=0.43]{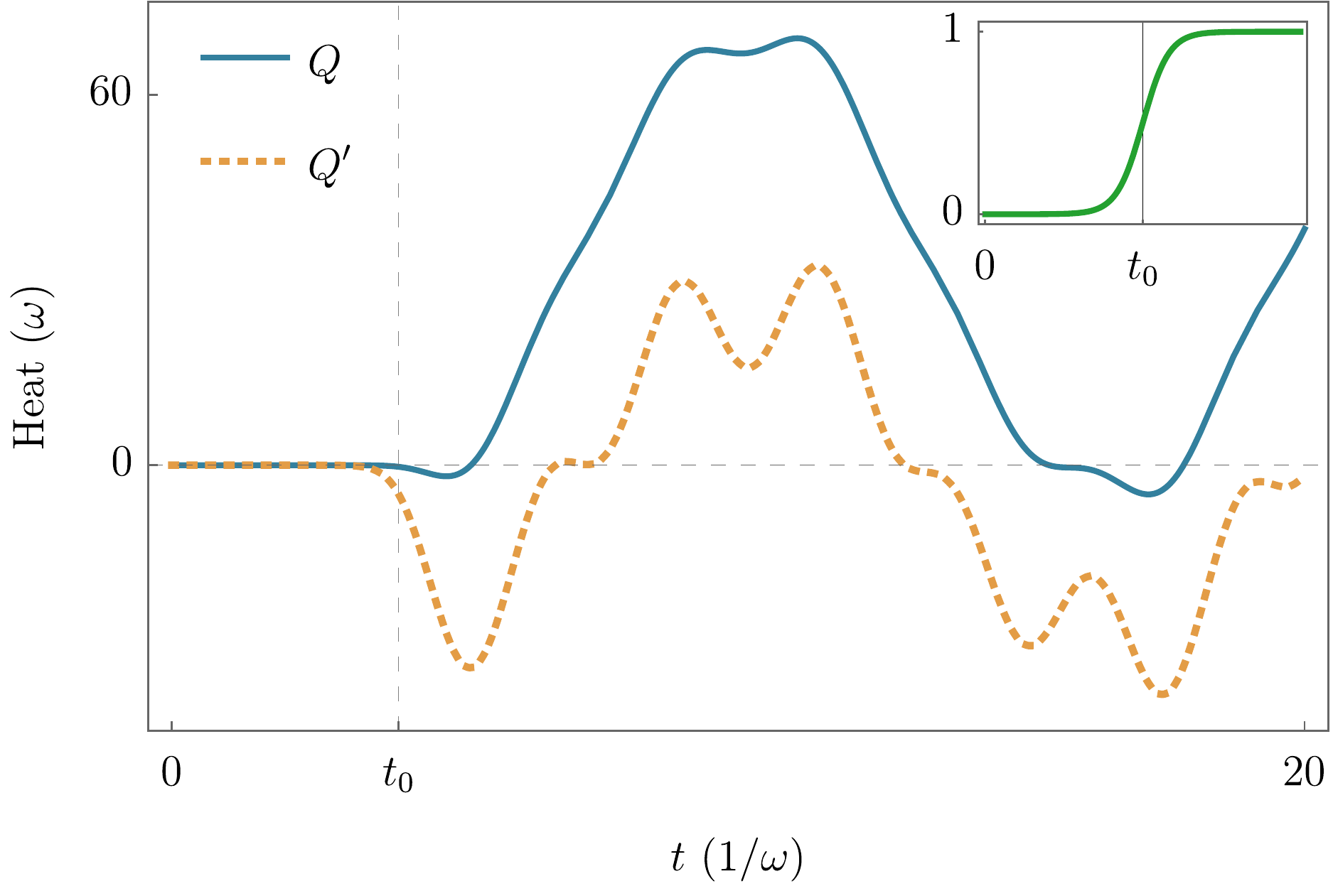}
\caption{
The naive heat $Q$ and the corrected heat $Q'$ are plotted with time. The inset shows the switching function $q(t)/q$. We have chosen $s$ to provide a smooth switch, which has midpoint $t_0$. We have chosen $\omega/\omega_{\rm m}=1$, $\eta=1/2$, and $\gamma =2\beta$ such that $Q$ tends to be positive despite an initial transient period. The corrected heat $Q'$ is significantly different to $Q$ in this coupling regime (ultrastrong) with $\delta Q$ providing an overall negative shift and additional oscillations.} 
\label{delQ}
\vspace*{2mm}
\end{center}
\end{minipage}
\end{figure}

We finally provide an example in which the non-trivial correction $\delta Q$ can be calculated explicitly, showing that it is not generally negligible in practical applications. We return to the dipole-field example of Sec.~\ref{noncon}. Although a heat bath is usually defined as consisting of infinitely many modes, we restrict our attention to a single photonic mode in a volume $v$ for simplicity. The extension to the multi-mode case is straightforward.  We further assume dipole canonical operators ${\bf r}=r{\bm \epsilon}$ and ${\bf p}=p{\bm \epsilon}$ where ${\bm \epsilon}$ is the mode polarisation vector. Allowing a time-dependent coupling parameter $q(t)$ gives the Hamiltonian $H=E_{A'}(t)+E_B$ where
\begin{align}
&E_{A'}(t) ={1\over 2}m{\dot r}(t)^2+V(r)= {1\over 2m}(p-q(t)A)^2+V(r),\\
&E_B = H_B=\omega a^\dagger a,
\end{align}
in which $A=(a^\dagger +a)/\sqrt{2\omega v}$ and $m{\dot r}(t):=p-q(t)A$. The transformation in Eq.~(\ref{pzw}) between frames $X$ and $Y$ becomes
\begin{align}
R(t)=\exp\left(-iq(t)rA\right).
\end{align}
The correction $\delta H_B(t)$ is easily found to be
\begin{align}
\delta H_B(t) = q(t)r \Pi - {q(t)^2 r^2\over 2v}
\end{align}
where $\Pi = i\sqrt{\omega/(2v)}(a^\dagger -a)$ is conjugate to $A$ in the sense that $[A,\Pi]=i/v$. In frame $X$ the operator $\Pi$ represents the observable $-E_{\rm T} = {\dot A}=-i[A,H(t)]=\Pi$. Thus, in frame $Y$ the observable $-E_{\rm T}$ is explicitly time-dependent and the operator $\Pi$ represents the transverse Maxwell displacement $-D_{\rm T} = -E_{\rm T}(t) - P_{\rm T}(t)$ where $P_{\rm T}(t) = q(t)r/v$ is the dipole's polarisation.

To go further we now assume the potential $V(r)= m \omega_{\rm m}^2 r^2/2$. The composite therefore consists of interacting quantum harmonic oscillators with time-dependent coupling function $q(t)$. We assume a function that smoothly increases from $0$ to $1$ near a point $t_0$, given by
\begin{align}
{q(t)\over q} = {1\over 2}\left(1+ \tanh \left[{t-t_0\over s}\right]\right)
\end{align}
where parameter $s$ controls the smoothness of the switch. We assume an initially uncorrelated state $\rho_A^{\rm eq}(\gamma)\otimes \rho_B^{\rm eq}(\beta)$ (at $t=0$) in which system and bath are at equilibrium with temperatures $\gamma^{-1}$ and $\beta^{-1}$ respectively and we define the dimensionless coupling-parameter $\eta = q/(\omega \sqrt{mv})$. Values $\eta\sim 0.01$ specify weak-coupling while values $\eta \geq 0.1$ specify ultrastrong-couplings and values $\eta\geq 1$ are called deep-strong couplings \cite{kockum_ultrastrong_2019}. Fig.~\ref{delQ} shows the correction $\delta Q$, which oscillates in time and with coupling strength. The oscillations become more pronounced with increasing coupling and can be significant for sufficiently strong coupling.

\section{Conclusions}\label{conc}

When the measurable energy of a subsystem includes the interaction Hamiltonian the standard definition of reduced subsystem state, as specified by a conjugate subsystem decomposition, cannot be employed. We have shown that this situation is indeed encountered in prevalent physical models and we have therefore provided an alternative subsystem decomposition, which we have termed non-conjugate. 

Energy and information changes in non-conjugate quantum subsystems possess significantly different behaviour. In particular, since operations performed on a subsystem $A'$ can disturb a distinct non-conjugate subsystem $B$, they can increase uncertainty about system $B$. We have proven both first and second laws of thermodynamics for non-conjugate subsystems, which generalises previous results \cite{esposito_entropy_2010,esposito_second_2011,deffner_information_2013,strasberg_quantum_2017,strasberg_first_2021}, while also ensuring internal consistency between the employed definitions of energy and entropy of the working system. Beyond these two basic laws, further additive laws for the Von Neumann and observational entropy changes of the measurable working system $A'$ and non-conjugate physical bath $B$ do not generally hold. This leads either to a breakdown of Clausius and free-energy (Landauer-type) inequalities that relate entropy and free energy changes of the working system to heat and work, or else to a non-trivial redefinition of heat and, in particular, work. More broadly, the notion of non-conjugate subsystems should provide a useful tool in understanding quantum interactions, helping to provide fundamental insight into the physical mechanisms underpinning energy and information exchanges.\\

\noindent {\em Acknowledgement.} I am indebted to P. Strasberg and A. Nazir, for comments on earlier drafts, and for useful discussions. I also thank P. Rangriz for useful discussions. This work was supported by the UK Engineering and Physical Sciences Research Council, grant no. EP/N008154/1.

\bibliography{non_conjugate.bib}

\onecolumngrid

\section*{Appendix: Material truncation}\label{qrm}

Often a truncation to the lowest two energy-levels of an atomic system (natural or artificial) is performed. Examples of paradigmatic models from light-matter physics are the well-known quantum Rabi model and the Jaynes-Cummings model, both of which describe the interaction of a TLS with a bosonic mode. A truncation of the bare energy $H_A$ in the dipole-mode model of Sec.~\ref{exa} results in a quantum Rabi model (QRM) Hamiltonian \cite{de_bernardis_breakdown_2018,stokes_gauge_2019,di_stefano_resolution_2019,roth_optimal_2019,stokes_gauge_2020}. However, the truncation results in non-equivalent models when performed in different frames \cite{de_bernardis_breakdown_2018,stokes_gauge_2019,di_stefano_resolution_2019,stokes_gauge_2020}. Pre-truncation the uniqueness of predictions found in different frames $X$ and $Y$ is guaranteed because ${\rm tr}(\rho_X O_X)={\rm tr}(\rho_Y O_Y)$ where $\rho_Y= R\rho_Y R^\dagger$ and $O_Y= RO_X R^\dagger$. However, by applying a truncating map $T$ one obtains truncated representations $T(O_X)$ and $T(O_Y)$ for which in general there does not exist a unitary operator $U$ acting on the truncated Hilbert such that $T(O_X)\neq UT(O_Y)U^\dagger$. In other words, under a given truncating map $T$, a frame $Z$ of the starting theory provides a  truncated theory $\tau_Z$, such that $\tau_Z$ and $\tau_{Z'}$ are not equivalent for $Z\neq Z'$. Uniqueness of predictions found using different $\tau_Z$ cannot therefore be guaranteed \cite{de_bernardis_breakdown_2018,stokes_gauge_2019,di_stefano_resolution_2019,stokes_gauge_2020}. Despite this, for a given purpose, a valid truncation can often be found via the comparison of post-truncated predictions with those obtained from the non-truncated theory \cite{de_bernardis_breakdown_2018,stokes_gauge_2019,di_stefano_resolution_2019,roth_optimal_2019,stokes_gauge_2020}. Once such a truncation has been identified, an equivalent model may of course be constructed subsequently by using a unitary operator $U$ defined over the truncated space.  

Here we consider the dipole-mode model of Sec.~\ref{exa}. To begin with we will assume that there are no external fields present and that the coupling parameter $q$ is time-independent, such that the Hamiltonian is time-independent in the Schr\"odinger picture. Let the zeroth and first eigenvalues of $H_A$ be denoted $\epsilon^0$ and $\epsilon^1$, and let $\omega_m:=\epsilon^1-\epsilon^0$ and $\Delta:=[\epsilon^1+\epsilon^0]/2$. Let ${\cal P}:=\ket{g}\bra{g}+\ket{e}\bra{e}$ project onto the corresponding two-level subspace. Truncation of $H_A$ may be defined as $H_A^2:={\cal P}H_A{\cal P} =\omega_m\sigma^z/2+\Delta$ where $\sigma^z := [\sigma^+,\sigma^-]$, $\sigma^+:=\ket{e}\bra{g}$, and $\sigma^- :=\ket{g}\bra{e}$. The standard truncated model in frame $X$ (Coulomb-gauge) is a QRM given by \cite{de_bernardis_breakdown_2018,stokes_gauge_2019,di_stefano_resolution_2019,roth_optimal_2019,stokes_gauge_2020}
\begin{align}\label{c2}
H_2= H_A^2+ \omega\left(a^\dagger a+{1\over 2}\right) + i{\tilde g}(a^\dagger +a)(\sigma^--\sigma^+)+\Omega(a+a^\dagger)^2
\end{align}
where $\Omega = q^2/(4m\omega v)$ and ${\tilde g} = d\omega_m/\sqrt{2\omega v}$ in which $d:=q\bra{\epsilon^0}r\ket{\epsilon^1}$ is assumed to be real. This is an expression of the energy within a frame $X_2$ of the truncated theory $\tau_X$. Similarly, the standard truncated model in frame $Y$ (multipolar-gauge) is a QRM given by \cite{de_bernardis_breakdown_2018,stokes_gauge_2019,di_stefano_resolution_2019,roth_optimal_2019,stokes_gauge_2020}
\begin{align}\label{m2}
H'_2= H_A^2+  {d^2\over 2v} + \omega\left(a^\dagger a+{1\over 2}\right) + ig(a^\dagger -a)(\sigma^++\sigma^-)
\end{align}
where $g= d\sqrt{\omega/(2v)}$. This is an expression in a frame $Y_2'$ of a {\em different} truncated theory $\tau_Y$. It is important to note that the two Hamiltonians $H_2$ and $H_2'$ are not merely expressions of the energy in different frames of the same theory, rather they are expressions of the energy provided by two distinct truncated theories. They are not unitarily equivalent in general. Nevertheless, one can define a unitary two-level model version of the rotation $R$ connecting frames $X$ and $Y$ of the non-truncated theory as \cite{di_stefano_resolution_2019}
\begin{align}\label{R2}
R_2 &= e^{-iq{\cal P}r{\cal P}A} = \exp\left(-i 
\eta\sigma^x[a^\dagger + a]\right) = \cos(\eta[a^\dagger + a])-i\sigma^x \sin(\eta [a^\dagger + a])
\end{align}
where $\eta := {g/ \omega}$. We can define a model $h_2' = R_2H_2 R_2^\dagger$, which is therefore expression of the energy within the truncated theory $\tau_X, $ with respect to a different frame, $Y_2$. Similarly, we can define a model $h_2= R_2^\dagger H_2' R_2$, which is an expression of the energy within the truncated theory $\tau_Y$ with respect to a different frame, $X'_2$. To first order in $q$ we have $R_2=I- i d \sigma_x A$ and with this approximation one finds that $h_2=H_2$ and $h_2'=H_2'$, such that the theories $\tau_X$ and $\tau_Y$ are seen to be equivalent up to first order in $q$ \cite{stokes_gauge_2020}. Their respective physical predictions may therefore be expected to be in good agreement within the weak-coupling regime. Regardless of the agreement between different truncated theories $\tau_Z$, a unitary transformation defined over the truncated space, such as $R_2$, connects frames within whatever truncated theory is under consideration and it can therefore be used to define non-conjugate subsystems therein. 

We now provide another example of non-conjugate subsystems arising within a well-known physical model, namely the QRM obtained via truncation in frame $Y$, which for a sufficiently anharmonic atomic system remains generally valid well into the ultrastrong-coupling regime \cite{de_bernardis_breakdown_2018,stokes_gauge_2019,di_stefano_resolution_2019,roth_optimal_2019}.
We note briefly that for atomic systems with lower anharmonicity, truncation can remain accurate for low energy predictions but it is no longer optimal to perform the truncation in frame $Y$ \cite{stokes_gauge_2019}. This is also the case for weaker couplings when considering multiple bosonic modes \cite{roth_optimal_2019}.

The total energy is defined as the sum of material kinetic and potential energies, plus the energy of the transverse electromagnetic field
\begin{align}
&H'(t) = E_{A'}(t)+E_B(t),\\
&E_{A'}(t)=H_A(t) = {p^2\over 2m}+\theta(r,t) = {1\over 2}m{\dot r}^2 +\theta(r,t),\\
&E_B(t)={v\over 2}\left([\Pi+P_{\rm T}(t)]^2+\omega^2 A^2\right)  = {v\over 2}\left(E_{\rm T}(t)^2+\omega^2 A^2\right) = H_B+V'(t)
\end{align}
where $p=m{\dot r}$, $P_{\rm T}(t)=q(t)r/v$, and $-E_{\rm T}(t) = {\dot A}(t) = \Pi+P_{\rm T}(t)$ is the transverse electric field associated with the single mode system. We have allowed a time-dependent coupling function $q(t)$ as well as an external potential that gives a time-dependent total potential $\theta(r,t)$. Truncation of $E_B(t)$ may be defined as \cite{de_bernardis_breakdown_2018,stokes_gauge_2019,di_stefano_resolution_2019}
\begin{align}
E_B^2(t) = {v\over 2}\left([\Pi+P^2_{\rm T}(t)]^2+\omega^2 A^2\right) = H_B+V'^2(t)
\end{align}
where $P^2_{\rm T}(t) = {\cal P}(t)P_{\rm T}(t){\cal P}(t) = d(t)\sigma^x(t)/v$ in which $\sigma^x(t):=\sigma^+(t)+\sigma^-(t)$ and $d(t) := q(t)\bra{g(t)}r\ket{e(t)}$ is assumed to be real. This results in the quantum Rabi Hamiltonian \cite{de_bernardis_breakdown_2018,stokes_gauge_2019,di_stefano_resolution_2019}
\begin{align}
H'_2(t) =& H_A^2(t)+ E_B^2(t) ={\omega_m(t)\over 2}\sigma^z(t) +\Delta(t) +\omega \left(a^\dagger a+{1\over 2}\right) + ig(t)[\sigma^+(t)+\sigma^-(t)](a^\dagger-a)+k(t)
\end{align}
where $g(t):= d(t)\sqrt{\omega/(2v)}$, and $k(t):=d(t)^2/(2v)$. A Jaynes-Cummings Hamiltonian is obtained if the so-called counter-rotating terms $\sim \sigma^+(t)a^\dagger$ and $\sim \sigma^-(t)a$ are neglected. This rotating-wave approximation is generally only valid in the weak-coupling regime and sufficiently close to resonance, i.e., when $g(t)\ll \omega_m(t)\approx \omega$. The physical TLS is a subsystem $A'$ with observables represented by operators of the form $O_A^2\otimes I_B$ in frame $Y'_2$. In particular the TLS energy is $E_{A'}^2(t)=H_A^2(t)$, the state of the TLS $A'$ is represented at time $t$ by $\rho_{A'}(t)={\rm tr}_B\rho'(t)$ where $\rho'(t)$ represents the state of the composite at time $t$ in frame $Y_2'$.

The time dependent generalisation of $R_2$ in Eq.~(\ref{R2}) is
\begin{align}\label{R2t}
R_2(t)= \exp\left(-i 
\eta(t)\sigma^x(t)[a^\dagger + a]\right) = \cos(\eta(t)[a^\dagger + a])-i\sigma^x(t) \sin(\eta(t) [a^\dagger + a])
\end{align}
and using $R_2(t)^\dagger [\Pi +P^2_{\rm T}(t)] R(t) = \Pi$ we obtain
\begin{align}
R_2(t)^\dagger [H_B+V'^2(t)] R_2(t) = \omega\left(a^\dagger a+{1\over 2}\right) =:H_B,
\end{align}
which represents the energy $E_B^2(t)$ in frame $X'_2$. The physical photonic mode is a subsystem $B$ with observables represented in frame $X'_2$ by operators of the form $I_A\otimes O_B^2$. The state of $B$ at time $t$ is $\rho_B(t)={\rm tr}\rho(t)$ where $\rho(t):=R_2(t)^\dagger \rho'(t)R_2(t)$ represents the state of the composite in frame $X'_2$ at time $t$. The energy $E^2_A(t)$ of the two-level system $A$ is represented in frame $X'_2$ by \cite{di_stefano_resolution_2019}
\begin{align}\label{ea2x}
&E_{A'}^2(t)=R_2(t)^\dagger H_A^2(t)R_2(t) =\Delta(t)+{\omega_m(t)\over 2} \left(\sigma^z(t)\cos[2\eta(t)(a^\dagger+a)] + \sigma^y(t)\sin[2\eta(t)(a^\dagger +a)]\right)
\end{align}
where $\sigma^y(t) := i[\sigma^+(t)-\sigma^-(t)]$. In frame $X'_2$ the two-level system energy clearly possesses a highly non-trivial dependence on the bosonic mode operators. In the case of a time-independent Hamiltonian, $H'^2(t)\equiv H'^2$, the rotation $R^2$ is also time-independent and the Hamiltonian in frame $X'_2$ becomes $H^2 = E_{A'}^2+H_B$ where $E_{A'}^2$ is given by Eq.~(\ref{ea2x}), but with all time-dependence dropped. This result is the same rotated two-level model energy presented in Ref.~\cite{di_stefano_resolution_2019} as an equivalent alternative Hamiltonian to the quantum Rabi Hamiltonian obtained via truncation in frame $Y'_2$.

\subsection{Example: Jaynes-Cummings model}\label{JC}

\begin{figure}[t]
\begin{minipage}{\columnwidth}
\begin{center}
\vspace*{-10mm}
\textbf{(a)}\hspace*{-1mm}\includegraphics[scale=0.45]{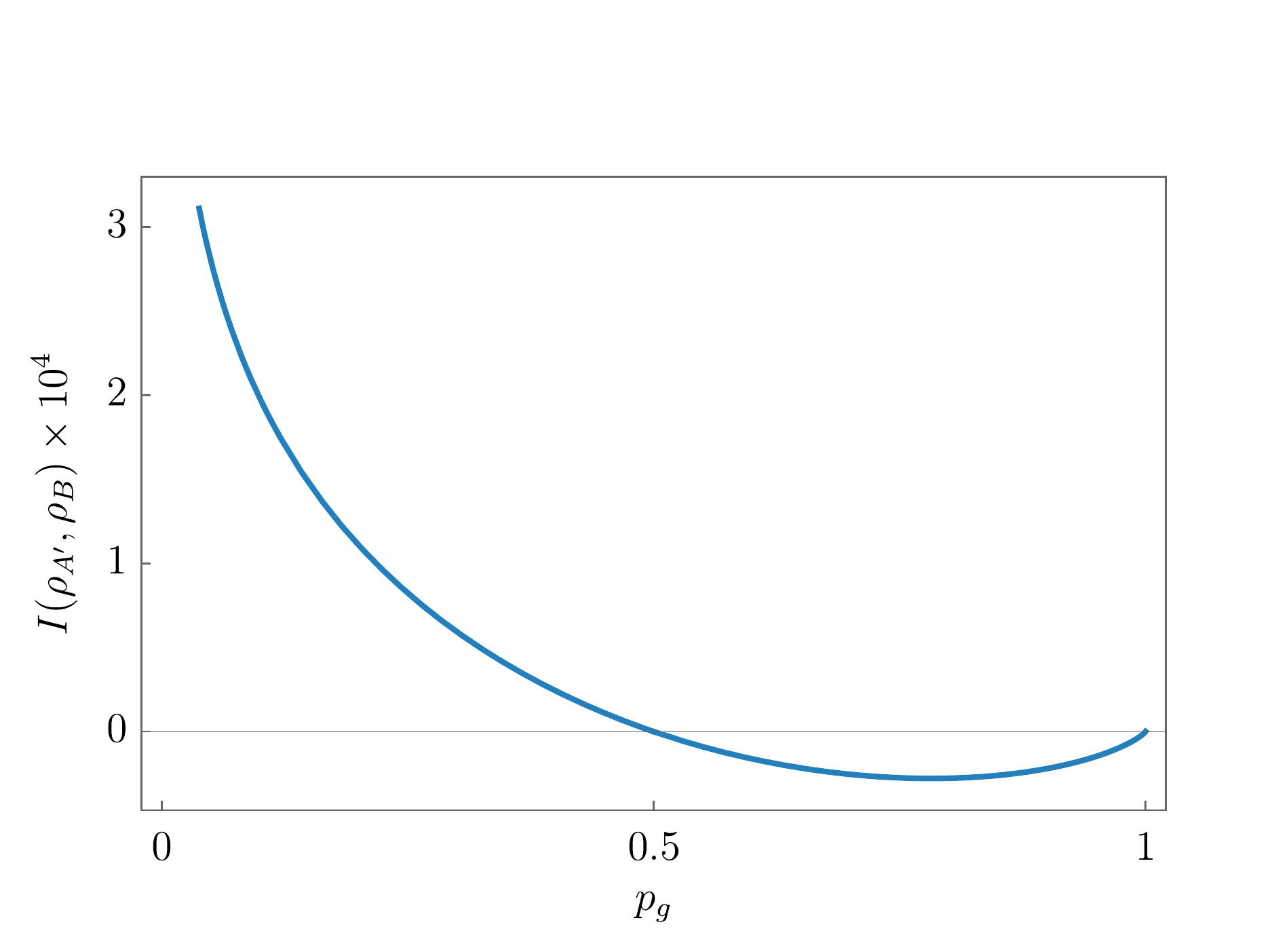}
\textbf{(b)}\hspace*{-1mm}\includegraphics[scale=0.45]{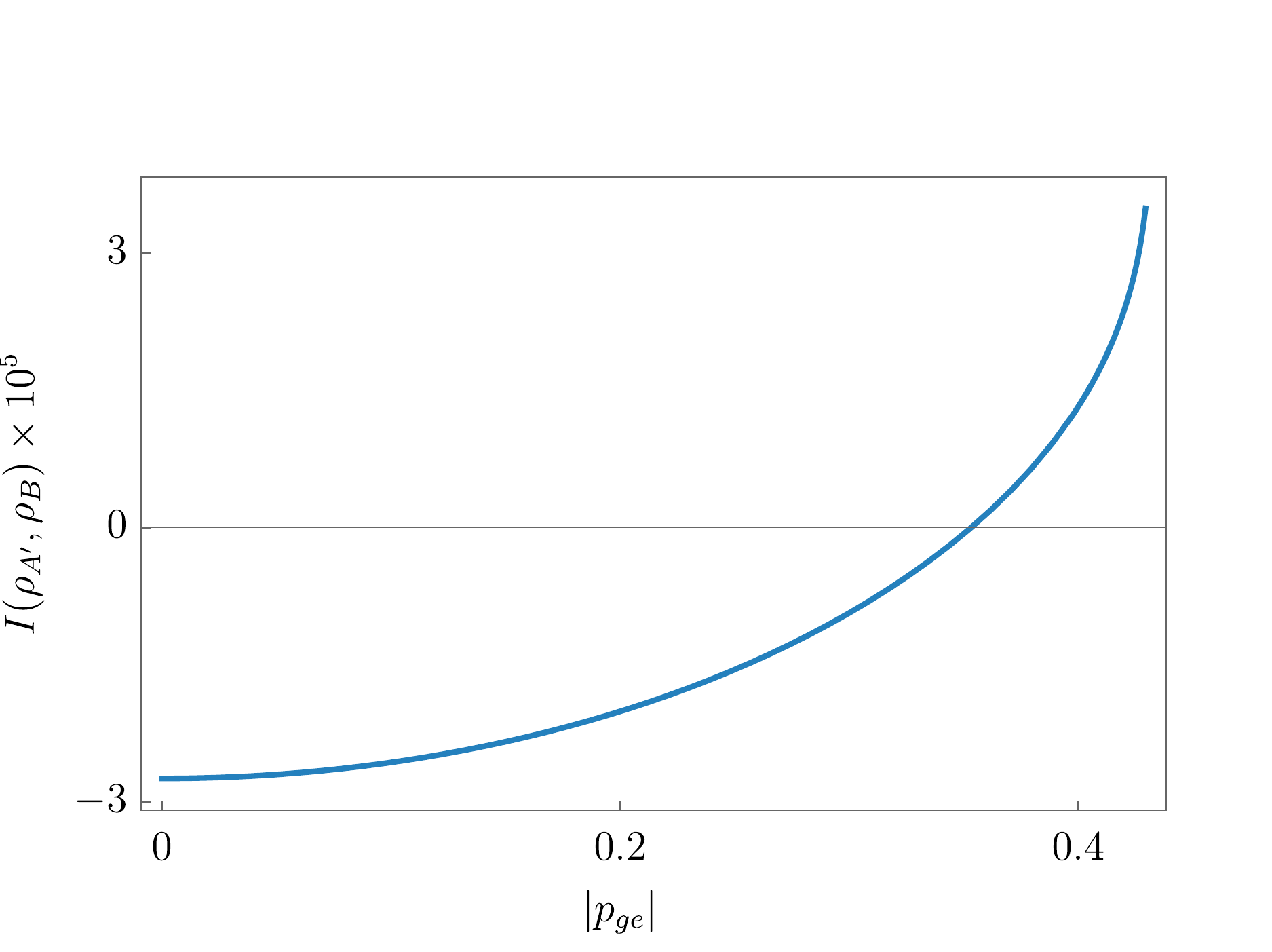}
\caption{\textbf{(a)} The mutual information of $A'$ and $B$ is plotted with $p_g$ assuming $\eta=0.01$ (weak-coupling) and $p_{ge}=0$. \textbf{(b)} The mutual information of $A'$ and $B$ is plotted with $|p_{ge}|$ assuming $\eta=0.01$ and $p_g=3/4$.}\label{MI_noncon}
\vspace*{-2mm}
\end{center}
\end{minipage}
\end{figure}

As an illustrative example of the entropies of non-conjugate subsystems, we consider a simplified version of the dipole-mode model of Sec.~\ref{exa}. The truncation of this model to include only two dipole energy levels $\{\ket{g},\ket{e}\}$ is discussed in detail in appendix \ref{qrm}. Here we consider the regime of weak and resonant interaction between the dipole and mode, without any external control. We perform the truncation in frame $X$ and neglect all non-resonant interactions to give the time-independent Jaynes-Cummings Hamiltonian
\begin{align}
H = \omega\left(a^\dagger a + {\sigma^z\over 2}\right) + ig\left(a^\dagger\sigma^--a\sigma^+\right)
\end{align}
where $\sigma^+ = \ket{e}\bra{g}$, $\sigma^-=(\sigma^+)^\dagger$, $\sigma^z=[\sigma^+,\sigma^-]$, $g=d\sqrt{\omega/(2v)}$, and $d=q\bra{e}r\ket{g}$ is assumed to be real. The energies of the two-level dipole are $\epsilon_g = -\omega/2$ and $\epsilon_e = \omega/2$. The transformation $R$ that connects frames $X$ and $Y$ has generator $G=-qrA$. By similarly truncating this operator and neglecting non-resonant terms we obtain ${\cal G} =-\eta(a^\dagger \sigma^- +a \sigma^+)$ where $\eta=g/\omega$. We note that the weak-coupling regime is defined by $\eta \sim 0.01$. We can define a truncated yet unitary version of the operator $R$ as
\begin{align}
{\cal R} := e^{i{\cal G}}= \exp\left(-i\eta[a^\dagger \sigma^+ +a\sigma^-] \right),
\end{align}
which connects to a new frame $Y$ within the Jaynes-Cummings theory. Using these definitions we can compute the physical dipole density operator $\rho_{A'}$ and associated quantities for a given $\rho$ representing the state of the composite system with respect to frame $X$. Suppose that the physical mode is prepared in the vacuum state such that $\rho$ is given by $\rho = \rho_A \otimes \ket{0_B}\bra{0_B}$. We obtain
\begin{align}
\rho_{A'} = K_0\rho_A K_0^\dagger + K_1 \rho_A K_1^\dagger
\end{align}
where
\begin{align}
K_0 &= \ket{g}\bra{g} + \ket{e}\bra{e} \cos \eta, \\
K_1 &= -i\ket{g}\bra{e}\sin \eta
\end{align}
are Krauss operators defining a quantum operation $\rho_A \to \rho_{A'}$. The populations and coherences $p'_g,\,p_{eg}'$ of $\rho_{A'}$, are related to the elements $p_g,\, p_{eg}$ of $\rho_A$ by
\begin{align}
p'_g &= p_g+(1-p_g)\sin^2 \eta, \\
p_{eg}' & = p_{eg} \cos \eta.
\end{align}
The remaining elements are $p_e=1-p_g$, $p_e'=1-p_g'$, $p_{ge} = p_{eg}^*$ and $p_{ge}'=p_{eg}'^*$. The quantity $I(\rho_{A'},\rho_B) = S(\rho_{A'})-S(\rho_A)$ is shown in Fig.~\ref{MI_noncon}. It can take negative values for given matrix elements $p_g,\, p_{eg}$, or equivalently, for given elements $p_g',\,p'_{eg}$. 

\end{document}